\documentclass[sn-mathphys,Numbered,iicol]{sn-jnl}

\usepackage{graphicx}%
\usepackage{multirow}%
\usepackage{amsmath,amssymb,amsfonts}%
\usepackage{amsthm}%
\usepackage{mathrsfs}%
\usepackage[title]{appendix}%
\usepackage[dvipsnames]{xcolor}%
\usepackage{textcomp}%
\usepackage{manyfoot}%
\usepackage{booktabs}%
\usepackage{algorithm}%
\usepackage{algorithmicx}%
\usepackage{algpseudocode}%
\usepackage{listings}%

\usepackage{units}
\usepackage{stmaryrd}
\usepackage{graphicx}
\usepackage[T2A,T1]{fontenc}
\usepackage{academicons}
\allowdisplaybreaks

\definecolor{orcidlogocol}{HTML}{A6CE39}

\allowdisplaybreaks
\usepackage{prettyref}
\newrefformat{sec}{Sect.~\ref{#1}}
\newrefformat{subsec}{Sect.~\ref{#1}}

\DeclareRobustCommand{\cyrtext}{%
  \fontencoding{T2A}\selectfont\def\encodingdefault{T2A}}
\DeclareRobustCommand{\textcyr}[1]{\leavevmode{\cyrtext #1}}

\usepackage{scalerel}
\usepackage{tikz}
\usetikzlibrary{svg.path}

\definecolor{orcidlogocol}{HTML}{A6CE39}
\tikzset{
  orcidlogo/.pic={
    \fill[orcidlogocol] svg{M256,128c0,70.7-57.3,128-128,128C57.3,256,0,198.7,0,128C0,57.3,57.3,0,128,0C198.7,0,256,57.3,256,128z};
    \fill[white] svg{M86.3,186.2H70.9V79.1h15.4v48.4V186.2z}
                 svg{M108.9,79.1h41.6c39.6,0,57,28.3,57,53.6c0,27.5-21.5,53.6-56.8,53.6h-41.8V79.1z M124.3,172.4h24.5c34.9,0,42.9-26.5,42.9-39.7c0-21.5-13.7-39.7-43.7-39.7h-23.7V172.4z}
                 svg{M88.7,56.8c0,5.5-4.5,10.1-10.1,10.1c-5.6,0-10.1-4.6-10.1-10.1c0-5.6,4.5-10.1,10.1-10.1C84.2,46.7,88.7,51.3,88.7,56.8z};
  }
}

\newcommand\orcidicon[1]{\href{https://orcid.org/#1}{\mbox{\scalerel*{
\begin{tikzpicture}[yscale=-1,transform shape]
\pic{orcidlogo};
\end{tikzpicture}
}{|}}}}

\begin{document}

\title[Signature of $f(R)$ gravity via Lema$\^$itre-Tolman-Bondi inhomogeneous perturbations]{\textbf{Signature of $f(R)$ gravity via Lemaître-Tolman-Bondi inhomogeneous perturbations}}


\author*[1,2]{\fnm{\textbf{Tiziano}} \sur{\textbf{Schiavone}} \orcidicon{0000-0003-0569-9570}} \email{tschiavone@fc.ul.pt} 

\author[3,4]{\fnm{\textbf{Giovanni}} \sur{\textbf{Montani}} \orcidicon{0000-0002-2550-5553}}


\affil[1]{\orgname{Galileo Galilei Institute for Theoretical Physics}, \orgaddress{\street{Largo Enrico Fermi 2}, \city{Florence}, \postcode{I-50125}, \country{Italy}}}

\affil[2]{\orgdiv{Instituto de Astrofisíca e Ciências do Espaço}, \orgname{Faculdade de Ciências da Universidade de Lisboa}, \orgaddress{\street{Edificio C8, Campo Grande}, \city{Lisbon}, \postcode{P-1740-016}, \country{Portugal}}}

\affil[3]{\orgdiv{Fusion and Nuclear Safety Department C.~R. Frascati}, \orgname{ENEA}, \orgaddress{\street{Via E. Fermi 45}, \city{Frascati (RM)}, \postcode{I-00044}, \country{Italy}}}

\affil[4]{\orgdiv{Physics Department}, \orgname{Sapienza University of Rome}, \orgaddress{\street{Piazzale A. Moro 5}, \city{Rome}, \postcode{I-00185}, \country{Italy}}}


\abstract{We analyze inhomogeneous cosmological models in the local Universe, described by the Lema\^itre-Tolman-Bondi (LTB) metric and developed using linear perturbation theory on a homogeneous and isotropic Universe background. Focusing on the different evolution of spherical symmetric inhomogeneities, we compare the $\Lambda$LTB model, in which the cosmological constant $\Lambda$ is included in the LTB formalism, with inhomogeneous cosmological models based on $f\left(R\right)$ modified gravity theories viewed in the Jordan frame. We solve the system of field equations for both inhomogeneous cosmological models adopting the method of separation of variables: we integrate analytically the radial profiles of local perturbations, while their time evolution requires a numerical approach. The main result of the analysis concerns the different radial profiles of local inhomogeneities due to the presence of a non-minimally coupled scalar field in the Jordan frame of $f\left(R\right)$ gravity. While radial perturbations follow a power-law in the $\Lambda$LTB model, Yukawa-like contributions appear in the $f\left(R\right)$ theory. Interestingly, this latter peculiar behavior of radial profile is not affected by the choice of the $f\left(R\right)$ functional form. The numerical solution of time-dependent perturbations exhibits a non-diverging profile. This work suggests that investigations about local inhomogeneities in the late Universe may allow us to discriminate if the present cosmic acceleration is caused by a cosmological constant term or a modified gravity effect.}


\maketitle

\clearpage

\tableofcontents{}

\section{Introduction}\label{intro}

Our current theoretical understanding in cosmology is essentially based on two crucial pillars: 1) General Relativity (GR) is the underlying gravitational theory; 2) the cosmological principle states that the spatial distribution of matter in the Universe is revealed as homogeneous and isotropic on scales sufficiently large. This paper aims to investigate the robustness of these two pillars, examining also possible deviations. 

Presently, the cosmological concordance model is generally referred to as the well-known $\Lambda$CDM model \cite{SupernovaSearchTeam:1998fmf,SupernovaCosmologyProject:1998vns,Weinberg:2008zzc,Montani-book2009hju}, which involves a cold dark matter (CDM) component and a cosmological constant $\Lambda$. This cosmological paradigm is consistent with most of the data, providing a reliable picture of the present-day observed Universe. For instance, maps of the cosmic microwave background radiation (CMB) \cite{Planck:2018vyg} have corroborated the idea of a homogeneous and isotropic Universe on large scales\footnote{This fact is widely accepted, although CMB anomalies and anisotropies have been recently emerged in the form of an observed temperature dipole \cite{Schwarz:2015cma}.}. The geometry of a purely homogeneous and isotropic Universe is properly described within the framework of the $\Lambda$CDM model through the Friedmann-Lema\^itre-Robertson-Walker (FLRW) metric \cite{Weinberg:2008zzc}.

Concerning the first pillar of the $\Lambda$CDM model, despite all the remarkable predictions and successes to explain theoretically many observational facts in the Universe, it is believed that GR is not the ultimate theory of gravity. Indeed, some open problems in cosmology, such as the need for dark components such as CDM and dark energy, the exact nature of which is still unknown, the cosmological constant problem \cite{Weinberg:1988cp, Peebles:2002gy}, and the Hubble constant tension\footnote{In particular, CMB measurements \cite{Planck:2018vyg} have provided a value of the actual expansion rate, \textit{i.e.}, the Hubble constant, that is incompatible with observations of local probes, such as Cepheids and type Ia supernovae (SNe Ia) \cite{pantheon+}, with a significance level of $4.9\,\sigma$. See \cite{Vagnozzi:2019ezj,DiValentino:2020zio,DiValentino:2021izs,Abdalla:2022yfr,Perivolaropoulos:2021jda,Vagnozzi:2023nrq} for comprehensive reviews. See also \cite{Lenart:2022nip,Bargiacchi:2023jse,Dainotti:2023ebr,Dainotti:2023bwq,Dainotti:2023fha,Dainotti:2023yrk,Montani:2023ywn,Bargiacchi:2023rfd,Dainotti:2024gca} for attempts to reduce the tension, using new statistical analysis and combining different cosmological probes.}, may be regarded as a possible signal of the breakdown of GR on galactic and cosmological scales and have motivated people to discuss some modifications of the theory. 

Among all the possible proposals of modified gravity with respect to the Einsteinian formulation, the so-called $f\left(R\right)$ gravity theories \cite{Buchdahl:1970ynr,Nojiri-Odintsov2007,Nojiri:2010wj-unified,Sotiriou-Faraoni2010,Faraoni:2010pgm,Tsujikawa:2010zza,DeFelice:2010aj,Capozziello:2011et,Nojiri:2017ncd-nutshell} stand for their simple morphology since de facto only an extra scalar degree of freedom is added to the gravitational field dynamics. These theories represent an extension of GR, a particular class of modified gravity, in which the scalar curvature $R$ in the gravitational Lagrangian density is replaced by a function $f\left(R\right)$, \textit{i.e.}, an extra geometrical degree of freedom with respect to GR. Moreover, scalar-tensor gravity provides an equivalent representation of $f\left(R\right)$ theories: adopting a formulation in the so-called Jordan frame \cite{Nojiri-Odintsov2007,Tsujikawa:2010zza,Sotiriou-Faraoni2010,Faraoni:2010pgm,DeFelice:2010aj,Nojiri:2017ncd-nutshell}, the additional mode is explicitly translated into a non-minimally coupled scalar field to standard gravity\footnote{As typical in scalar-tensor theories, the $f\left(R\right)$ gravity can be also described upon a proper conformal transformation in the so-called Einstein frame \cite{Nojiri-Odintsov2007,Tsujikawa:2010zza,Sotiriou-Faraoni2010,Faraoni:2010pgm,Nojiri:2017ncd-nutshell}, which is mathematically equivalent to the Jordan one. Actually, the problem of identifying the physical frame has been long debated \cite{Magnano:1993bd,Capozziello:1996xg,Capozziello:2006dj,Capozziello:2014bqa,Bahamonde:2016wmz,Bahamonde:2017kbs}. In this work, we decide to adopt the Jordan frame, since it is provided only by a pure redefinition of fields starting from the $f\left(R\right)$ metric formalism.}.

The interest in the $f\left(R\right)$ modified gravity is because extra degrees of freedom may allow us to find alternative explanations for the abovementioned unresolved problems in cosmology. For instance, some successful proposals in the $f\left(R\right)$ gravity \cite{Hu:2007nk,Song:2006ej,Starobinsky:2007hu,Tsujikawa:2007xu} describe the present accelerating Universe, avoiding to introduce ad hoc extra components, such as dark energy. Other examples of $f\left(R\right)$ models have been developed to try to alleviate or solve the Hubble constant tension \cite{Odintsov:2020qzd,DiValentino:2021izs,dainottiApJ-H0(z),DainottiGalaxies-H0(z),Schiavone:2022shz,Nojiri:2022ski,Schiavone:2022wvq,Montani:2023xpd}. Moreover, the presence of extra degrees of freedom within the $f\left(R\right)$ metric formalism is also discussed in gravitational-wave physics, regarding the nature of their polarizations \cite{Moretti:2019yhs}. 

Therefore, an increasing interest has risen in recent years to develop new methods to discriminate between the standard $\Lambda$CDM model and modified gravity cosmological scenarios for the present Universe. Extended $f\left(R\right)$ theories of gravity admit a larger number of solutions than Einstein field equations in GR, but extra degrees of freedom and non-linearity imply non-trivial cosmological dynamics. Some cosmological exact solutions in the $f\left(R\right)$ gravity have been found for simple scenarios, especially in a homogeneous and isotropic cosmology \cite{Saez-Gomez:2008rfn,Faraoni:2010pgm,Bisabr:2015jia,Ciftci:2017tjc,Faraoni:2021nhi}.

Concerning the second pillar of the $\Lambda$CDM model, that is the homogeneity and isotropy of the Universe on large scales, it should be emphasized that the cosmological principle is not invariant for any spatial scale: our present Universe seems to reach the conditions to be homogeneous on a scale of about $100\,\textrm{Mpc}$, as suggested by galaxy surveys and the large-scale structure of the Universe \cite{Yadav:2005vv,Sarkar:2009iga}. Thus, when considering physical phenomena that occur on smaller spatial scales, local features of the Universe lead inevitably to deviations from the FLRW geometry, which could affect cosmological parameters \cite{McClure:2007vv,Labini:2011tj,Kazantzidis:2020tko,Krishnan:2021dyb,Aluri:2022hzs} and the luminosity distance distribution \cite{Schiavone:2023olz}. See also \cite{Buchert:1999er,Gasperini:2009mu,Fanizza:2019pfp} for the development of the averaging formalism in cosmology to average scalar quantities in a limited region of space-time.

Furthermore, observational evidence of a local void (an underdense region) have emerged on scales of several hundreds of $\textrm{Mpc}$ \cite{Zehavi:1998gz,Keenan:2013mfa,Haslbauer:2020xaa,Wong:2021fvu}. In this regard, the Lema\^itre-Tolman-Bondi (LTB) model \cite{Lemaitre:1927zz,Tolman:1934za,Bondi:1947fta,Peebles:2002gy,Montani-book2009hju} is widely employed to describe spherically symmetric non-stationary inhomogeneities in the local Universe, while it approaches a homogeneous Universe far enough from the center of symmetry. The LTB solution with the presence of a cosmological constant $\Lambda$ is commonly referred to as the $\Lambda$LTB model, which has been studied for many decades as a possible (simplified) framework to consider local deviations of the Universe today from the homogeneity \cite{Garcia-Bellido:2008vdn,Sinclair:2010sb,Moss:2010jx,Fanizza:2014tua,Cosmai:2018nvx,Lukovic:2019ryg,Camarena:2021mjr} or to alleviate the Hubble constant tension \cite{Kenworthy:2019qwq,Ding:2019mmw,Cai:2020tpy,Castello:2021uad,Camarena:2022iae}. 

Since $f\left(R\right)$ theories and the $\Lambda$LTB model separately can only alleviate the Hubble constant tension, the combination of more than one non-standard physical effect, \textit{i.e.}, modified gravity and inhomogeneous cosmology, may be needed to accommodate these data inconsistencies into a cosmological model.

However, in this work, we do not focus on the Hubble tension, but we consider a prodromic question, \textit{i.e.}, the formulation of the LTB cosmology in modified gravity. Other papers were carried out to find inhomogeneous LTB solutions in extended gravity scenarios \cite{Harada:2001kc,Sharif:2014bta,Sussman:2017qya,Yu:2019cku,Bhatti:2021crd,Najera:2021aod,Najera:2022jvm}; here we consider the $f\left(R\right)$ metric formalism to highlight possible peculiarities of local inhomogeneities compared to the GR solutions. Our aim is to search for some specific markers of the matter distribution, which could allow us to distinguish a standard paradigm from an alternative cosmology. Indeed, local inhomogeneities incorporated in a modified gravitational scenario might become a predictive tool.

In this paper, in the spirit of describing small deviations from the homogeneity of the Universe today, we treat the inhomogeneous LTB metric as the flat FLRW background solution with the addition of small spherically symmetric perturbations, following a linear perturbation approach\footnote{Differently, in \cite{Clarkson:2007yp,Zibin:2008vj,Clarkson:2009sc} the linear perturbation theory is analyzed in the LTB cosmology considering the background metric as LTB to understand the evolution of perturbations on spherically symmetric spacetimes. Here, instead, we consider perturbative deviations from FLRW as the background metric to build exactly the LTB spacetime at the linear-order perturbation theory}. We extend the work carried out in \cite{Marcoccia:2018anj,Schiavone:2021tvh}. The present analysis aims to discriminate between the $\Lambda$LTB cosmological paradigm and the LTB solution as emerging in the Jordan frame of the $f\left(R\right)$ gravity, comparing the different evolution of inhomogeneous perturbations within these two schemes to describe the local behavior of the actual Universe. In this regard, we evaluate the role of a cosmological constant with respect to the presence of a non-minimally coupled scalar field in the Jordan frame, examining separately the cosmological dynamics. In particular, regarding the background cosmology in the Jordan frame, we adopt the $f\left(R\right)$ model developed by Hu and Sawicki \cite{Hu:2007nk}, which provides an interesting alternative to the dark energy component for the current cosmic acceleration within the $f\left(R\right)$ gravity.

Therefore, we investigate the first-order perturbation field equations for local inhomogeneities both in the $\Lambda$LTB and $f\left(R\right)$ inhomogeneous cosmological models. More specifically, the analysis of the 0-1 component of the gravitational field equations points out discrepancies between the two formalisms considered, due to the presence of the non-minimally coupled scalar field. In our analysis of the first-order perturbation equations, we adopt the separable variables method as a mathematical technique to address the solution of the partial differential equations system. Then, we separately obtain the time evolutions and radial profiles of perturbations within both two cosmological models. Actually, we get an analytic expression only for the radial part, while we need to numerically evaluate the time evolution of perturbations.

The main result of our work is the different radial patterns of inhomogeneous perturbations in GR and the $f\left(R\right)$ cosmology. Indeed, in the former case, we deal with a radial solution that is basically a power law; in the latter case, a peculiar Yukawa-like dependence emerges in the Jordan frame gravity whatever the $f\left(R\right)$ model considered. Both these radial solutions suggest that inhomogeneities decay rapidly on large scales according to the cosmological principle, but furthermore, from a theoretical point of view, we have a specific trace of how a modified gravity model can be distinguished from a standard gravity scenario in the presence of a cosmological constant. In other words, the Yukawa-like radial solution is a peculiar feature of the $f\left(R\right)$ gravity.

Regarding our numerical analysis of the time dependence of the obtained solutions, we show that local inhomogeneities are governed by a dynamics preserving the stability of the isotropic Universe both in the $\Lambda$LTB and modified gravity theories. Therefore, these stable-time solutions are physically acceptable and can be thought of as large-scale corrections to the FLRW geometry.

This work provides an interesting arena to investigate different scenarios of the clumpy inhomogeneous cosmology; at the same time, the method developed in this paper supplies a useful criterion to distinguish between GR and $f\left(R\right)$ modified gravity models on cosmological scales.

This paper is structured as follows: in \prettyref{sec:f(R)-modified-gravity-JF} we introduce the $f\left(R\right)$ modified theories of gravity in the Jordan frame and the Hu-Sawicki model; in \prettyref{sec: LTB in GR} and \prettyref{sec:LTB-JF} we implement the cosmological dynamics to the $\Lambda$LTB model in GR and the LTB spherically symmetric solution as emerging in the Jordan frame of the $f\left(R\right)$ gravity, respectively; in \prettyref{sec:pert-approach-GR} we show our perturbation approach to study local inhomogeneities in the $\Lambda$LTB model, obtaining background and first-order perturbation solutions, while in \prettyref{sec:Perturbation-approach-JF} we proceed similarly for the $f\left(R\right)$ theories in the Jordan frame; lastly, we summarize our results and conclusions in \prettyref{sec:conclusions}.

We adopt the metric signature $\left(-,+,+,+\right)$ throughout the paper, and we use natural units for the speed of light $c=1$. We denote with $\chi\equiv8\,\pi\,G$ the Einstein constant, being $G$ the gravitational Newton constant. The components of the gravitational field equations are sometimes denoted with a pair of indices $\mu-\nu$, in which the indices $\mu,\nu=0,1,2,3$ move along the set of coordinates.

\section{$f\left(R\right)$ modified gravity in the Jordan frame}\label{sec:f(R)-modified-gravity-JF}

We investigate the equivalence between $f\left(R\right)$ theories of gravity and the scalar-tensor representation in the Jordan frame \citep{Buchdahl:1970ynr,Nojiri-Odintsov2007,Tsujikawa:2010zza,Sotiriou-Faraoni2010,Nojiri:2010wj-unified,DeFelice:2010aj,Capozziello:2011et,Faraoni:2010pgm,Nojiri:2017ncd-nutshell}. There is no a priori motivation to consider a linear gravitational Lagrangian density with respect to the Ricci scalar $R$, apart from obtaining a second-order partial differential system for the field equations. 

The gravitational Lagrangian density, which is given by $R$ in GR in the Einstein-Hilbert action, is generalized in the context of $f\left(R\right)$ theories as a function $f$ of $R$, \textit{i.e.}, an extra degree of freedom. In addition, considering a matter term $S_{M}$, the total action of $f\left(R\right)$ gravity is given by
\begin{equation}
S=\frac{1}{2\,\chi}\,\int d^{4}x\,\sqrt{-g}\,f\left(R\right)+S_{M}\left(g_{\mu\nu},\psi\right)\,,\label{eq: total action f(R)}
\end{equation}
where $g$ is the determinant of the metric tensor with components $g_{\mu\nu}$, and $\psi$ refers to the matter fields. 

It can be shown that the extended gravitational field equations within the $f\left(R\right)$ metric formalism are fourth-order partial differential equations in the metric. If $f\left(R\right)=R$, specifically, the fourth-order terms vanish, and field equations reproduce exactly the Einstein field equations in GR. 

To bring field equations to a form that is easier to handle, the $f\left(R\right)$ gravity can be restated in the scalar-tensor formalism. In particular, in the so-called Jordan frame\footnote{It is quite similar to the prototype of extended gravitational theories, the Brans-Dicke formulation \citep{Brans:1961sx,Nordtvedt:1970uv} with a non-zero potential and a null Brans-Dicke parameter or the O'Hanlon proposal \citep{OHanlon:1972xqa}.}, it can be checked that the following action
\begin{equation}
S_{J}=\frac{1}{2\,\chi}\,\int_{\Omega}d^{4}x\,\sqrt{-g}\,\left[\phi\,R-V\left(\phi\right)\right]+S_{M}\label{eq: azione jordan frame}
\end{equation}
is dynamically equivalent to the $f\left(R\right)$ action given by Eq.~\eqref{eq: total action f(R)} if $f^{\prime\prime}\left(R\right)\neq0$, where the scalar field
\begin{equation}
\phi\equiv f^{\prime}\left(R\right)\label{eq: def scalar field}
\end{equation}
is governed by the scalar field potential defined as
\begin{equation}
V\left(\phi\right)\equiv\phi\,R\left(\phi\right)-f\left[R\left(\phi\right)\right]\,.\label{eq: def scalar field potential}
\end{equation}
It should be noted that the extra degree of freedom given by the $f\left(R\right)$ function turns into a scalar field\footnote{Actually, as shown in \citep{Olmo:2006eh}, the relation $f^{\prime\prime}\left(R\right)\neq0$ is a redundant requirement for the dynamic equivalence between an $f\left(R\right)$ theory and a scalar-tensor formulation in the Jordan frame. Indeed, it is sufficient to require that $f^{\prime}\left(R\right)$ be invertible (continuous and one-to-one in a given interval), i.e., the existence of $R=R\left(f^{\prime}\right)$. In this way, it is possible to build a scalar field potential $V\left(\phi\right)$.} $\phi$, which is non-minimally coupled to the metric. Conversely, there is only a minimal coupling for the matter.

Variations of the action \eqref{eq: azione jordan frame} with respect to the metric and scalar field lead to field equations in the Jordan frame
\begin{subequations}
\begin{align}
G_{\mu\nu}= & \frac{\chi}{\phi}\,T_{\mu\nu}-\frac{1}{2\,\phi}\,g_{\mu\nu}\,V\left(\phi\right)\nonumber\\
& +\frac{1}{\phi}\,\left(\nabla_{\mu}\nabla_{\nu}\phi-g_{\mu\nu}\,\boxempty\phi\right)\label{subeq: field equations Jordan frame - metric}\\
R= & \frac{dV}{d\phi},\label{subeq: field equations Jordan frame - scalar}
\end{align}
\end{subequations}
respectively, where $G_{\mu\nu}$ is the Einstein tensor, $T_{\mu\nu}$ is the stress-energy tensor of matter, $\nabla_{\mu}$ is the covariant derivative associated with the Levi-Civita connection of the metric, and $\boxempty=g^{\rho\sigma}\,\nabla_{\rho}\nabla_{\sigma}$. 

Furthermore, by taking the trace of Eq.~\eqref{subeq: field equations Jordan frame - metric} and using Eq.~\eqref{subeq: field equations Jordan frame - scalar}, a dynamical equation for the scalar field is obtained
\begin{equation}
3\,\boxempty\phi+2\,V\left(\phi\right)-\phi\,\frac{dV}{d\phi}=\chi\,T\label{eq: scalar field eq. Jordan frame}
\end{equation}
for a given matter source, where $T$ is the trace of the matter stress-energy tensor. 

Although the presence of a non-minimally coupled scalar field implies non-trivial dynamics in the Jordan frame, it is often convenient to adopt the Jordan frame of the $f\left(R\right)$ gravity, since the field equations \eqref{subeq: field equations Jordan frame - metric}, \eqref{subeq: field equations Jordan frame - scalar}, and \eqref{eq: scalar field eq. Jordan frame} are now second-order differential equations. 

Looking at the action given in Eq.~\eqref{eq: azione jordan frame}, the extra degree of freedom provided by $f\left(R\right)$ does not affect the matter action even in the Jordan frame. Therefore, the stress-energy tensor of ordinary matter must be divergence-free as in GR: $\nabla_{\nu}T^{\mu\nu}=0$. Since Bianchi identities are kept in modified gravity, the Einstein tensor is also covariant divergence-free: $\nabla_{\nu}G^{\mu\nu}=0$. Here, we define an effective stress-energy tensor related to the scalar field
\begin{equation}
T_{\mu\nu}^{[\phi]}=-\frac{1}{2\,\phi}\,g_{\mu\nu}\,V\left(\phi\right)+\frac{1}{\phi}\,\left(\nabla_{\mu}\nabla_{\nu}\phi-g_{\mu\nu}\,\boxempty\phi\right)\label{eq: def effective stress-energy tensor}
\end{equation}
to rewrite the field equations \eqref{subeq: field equations Jordan frame - metric} in the Jordan frame as
\begin{equation}
G_{\mu\nu}=\frac{\chi}{\phi}\,T_{\mu\nu}+T_{\mu\nu}^{[\phi]}\,.
\end{equation}

Finally, considering the divergence-free relations above, it is trivial to show that
\begin{equation}
\nabla_{\nu}T^{[\phi]\,\mu\nu}=\frac{\chi}{\phi^{2}}\,T^{\mu\nu}\,\nabla_{\nu}\phi\,.\label{eq: effective stress-energy tensor law}
\end{equation}
The effective stress-energy tensor $T_{\mu\nu}^{[\phi]}$ does not satisfy the usual law of the ordinary matter \citep{Santiago:1999by,Torres:2002pe,Faraoni:2004pi,Koivisto:2005yk}, unless in vacuum ($T_{\mu\nu}=0$). The extra scalar degree of freedom in the Jordan frame is quite different from a matter field; actually, $\phi$ is an effective scalar field originating from a scalar mode intrinsically due to a modification in the gravitational action. Note that the laws~\eqref{eq: effective stress-energy tensor law} describing the dynamics of $T_{\mu\nu}^{\left[\phi\right]}$, as well as the continuity equation related to $T_{\mu\nu}$, are not independent of the field Eqs.~\eqref{subeq: field equations Jordan frame - metric} and \eqref{eq: scalar field eq. Jordan frame}. However, Eqs.~\eqref{eq: effective stress-energy tensor law} can be employed as auxiliary equations to rewrite field equations in a different equivalent form. 

\subsection{The Hu-Sawicki model}\label{subsec:The-Hu-Sawicki-model}

Among several proposals for the functional form of the $f\left(R\right)$, one of the most studied dark energy models is provided by Hu and Sawicki \citep{Hu:2007nk,Song:2006ej}. It is useful to refer to the deviation $F\left(R\right)$ from the gravitational Lagrangian density in GR, i.e. 
\begin{equation}
f\left(R\right)=R+F\left(R\right)\,.\label{eq:f=00003DR+F}
\end{equation}

The functional form of the deviation $F\left(R\right)$ for the Hu-Sawicki (HS) model is
\begin{equation}
F\left(R\right)=-m^{2}\,\frac{c_{1}\,\left(\nicefrac{R}{m^{2}}\right)^{n}}{c_{2}\,\left(\nicefrac{R}{m^{2}}\right)^{n}+1}\,,\label{eq:F(R)-HS}
\end{equation}
where $n$ is a positive integer, $c_{1}$ and $c_{2}$ are the HS dimensionless parameters, and $m^{2}\equiv\chi\,\rho_{m0}/3$ with
$\rho_{m0}$ matter density today. 

Note that for $R\gg m^{2}$ the limiting case with an effective cosmological constant $\Lambda_{\textrm{eff}}=c_{1}\,m^{2}/2\,c_{2}$ is recovered. Approximating the cosmic accelerated phase of a flat $\Lambda$CDM model with an effective cosmological constant, we obtain the first constraint on the parameters $c_{1}$ and $c_{2}$:
\begin{equation}
\frac{c_{1}}{c_{2}}\approx6\,\frac{\Omega_{\Lambda0}}{\Omega_{m0}}\,,\label{eq:constraint1-c1c2}
\end{equation}
in which we have used the definitions of cosmological density parameters $\Omega_{m0}\equiv \rho_{m0}/\rho_{c0}$ and $\Omega_{\Lambda0}\equiv \rho_{\Lambda0}/\rho_{c0}$ for the matter component and the cosmological constant, respectively, being $\ensuremath{\rho_{c0}\equiv3H_{0}^{2}/\chi}$ the critical energy density of the Universe today, $\rho_{\Lambda 0}\equiv \Lambda/\chi$ the energy density associated with $\Lambda$, and $H_0$ is the Hubble constant. The subscript 0 denotes the present cosmic time $t_{0}$ at the redshift $z=0$.

Furthermore, Hu and Sawicki \citep{Hu:2007nk} have shown that today $R_{0}\gg m^{2}$, and also the approximation $R\gg m^{2}$ is viable for the entire past cosmic expansion. 

Hereinafter, we set $n=1$ for simplicity, focusing on the most extensively studied scenario in the HS gravity. In that case, the derivative $F_{R}\equiv dF/dR$ for $R\gg m^{2}$ is approximately
\begin{equation}
F_{R}\approx-\frac{c_{1}}{c_{2}^{2}}\,\left(\frac{m^{2}}{R}\right)^{2}\,.\label{eq:F_R-HS}
\end{equation}
In particular, for a flat $\Lambda$CDM model
\begin{equation}
\frac{R}{m^{2}}=3\,\left(\frac{1}{a^{3}}+4\,\frac{\Omega_{\Lambda0}}{\Omega_{m0}}\right)\,,\label{eq:R/m^2}
\end{equation}
where $a=a\left(t\right)$ is the scale factor in terms of the cosmic time. We have used the well-known Friedmann equations and the definition of the Ricci scalar $R$ in a flat FLRW metric \citep{Weinberg:2008zzc}. 

We can rewrite Eq.~\eqref{eq:F_R-HS} evaluated today in the limiting case for $R\gg m^{2}$:
\begin{equation}
F_{R0}\approx-\frac{c_{1}}{c_{2}^{2}}\,\left[3\,\left(1+4\,\frac{\Omega_{\Lambda0}}{\Omega_{m0}}\right)\right]^{-2}\,,\label{eq:FR0-condition-HS}
\end{equation}
in which we have implicitly assumed that the value of $\Lambda_{\textrm{eff}}$ to be the same as $\Lambda$ in the $\Lambda$CDM limit. We have also set the conventional notation for the scale factor today $a_{0}=1$. Hence, by setting a reference value for $F_{R0}$, we can obtain the second constraint on $c_{1}$ and $c_{2}$. Note that, according to Eqs.~\eqref{eq: def scalar field} and \eqref{eq:f=00003DR+F}, we have in the Jordan frame:
\begin{equation}
\phi=1-F_{R}\,.\label{eq:JF-deviation-GR}
\end{equation}
It should be noted that $F_{R}$ quantifies the deviation from the GR scenario, where $\phi=1$. 

The scalar field potential $V\left(\phi\right)$ in the Jordan frame for the HS model assumes the following form:
\begin{equation}
V\left(\phi\right)=\frac{m^{2}}{c_{2}}\,\left[c_{1}+1-\phi-2\sqrt{c_{1}\,\left(1-\phi\right)}\right]\,,\label{eq:scalar-field-potential-HS}
\end{equation}
where we used the definitions given in Eqs.~\eqref{eq: def scalar field} and \eqref{eq: def scalar field potential} referred to the $F\left(R\right)$ function in Eq.~\eqref{eq:F(R)-HS}. Moreover, we have selected the branch for the potential related to a minus sign just before the square root in Eq.~\eqref{eq:scalar-field-potential-HS} to converge to an asymptotically stable de Sitter Universe \citep{Saez-Gomez:2012uwp,delaCruz-Dombriz:2015tye}.

\begin{figure}
\centering \includegraphics[scale=0.24]{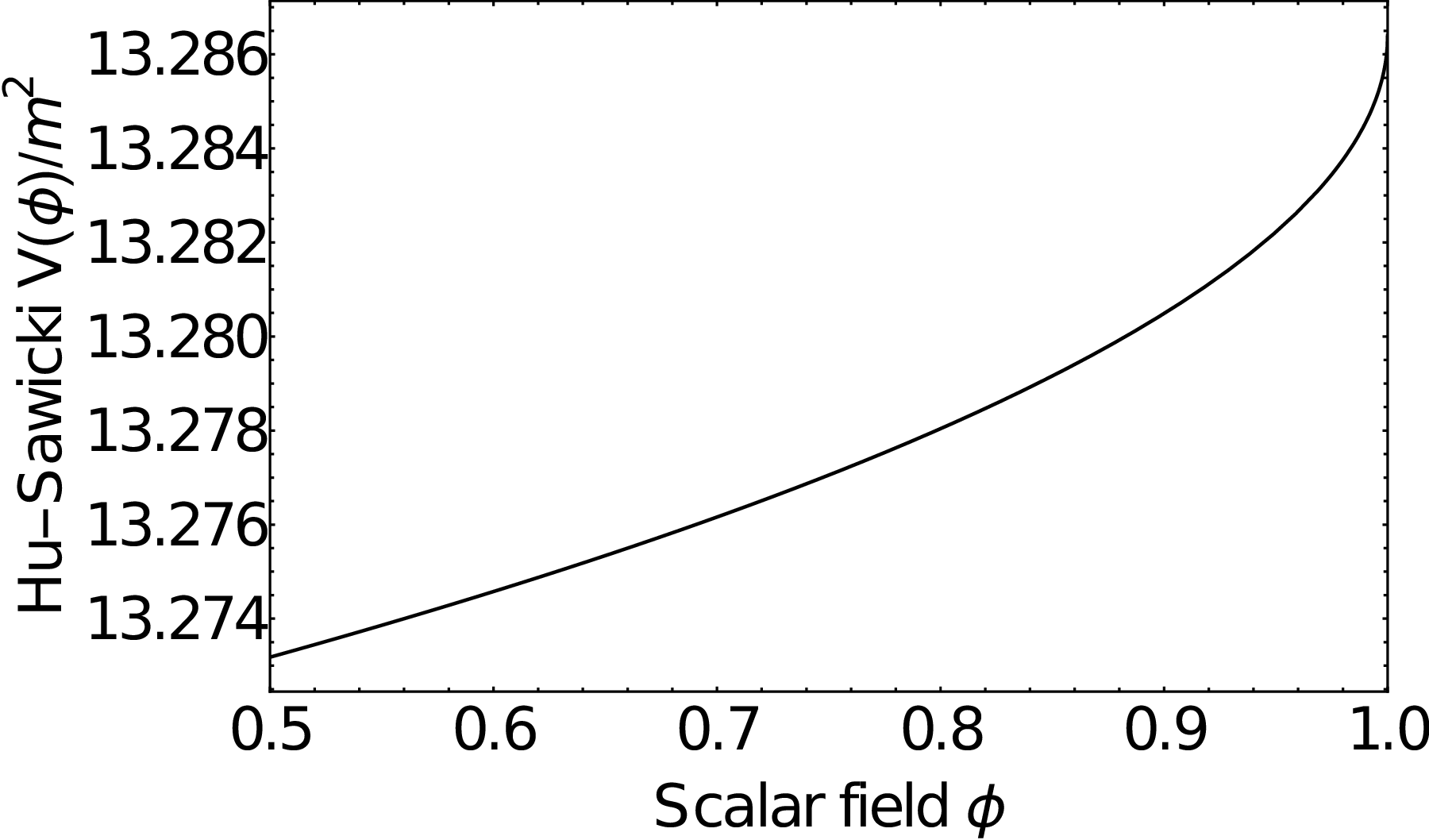}

\caption{Behavior of the HS scalar field potential in the Jordan frame, defined in Eq.~\eqref{eq:scalar-field-potential-HS}. It should be noted that $V\left(\phi\right)/m^{2}$ is dimensionless.}
\label{fig:HS-background-potential}
\end{figure}

Finally, we show two reasonable values for the HS dimensionless parameters $c_{1}$ and $c_{2}$. More precisely, we fix $\Omega_{m0}=0.3111$, and $\Omega_{\Lambda0}=0.6889$ from the Planck measurements \citep{Planck:2018vyg}; we set the value of the derivative of the field at the present cosmic time $\left|F_{R0}\right|=1.0\times10^{-7}$, considering the strongest bound between solar system \citep{Hu:2007nk} and cosmological constraints \citep{Lombriser:2014dua,Burrage:2017qrf,Liu:2017xef}. Then, using the conditions \eqref{eq:constraint1-c1c2} and \eqref{eq:FR0-condition-HS} for the $\Lambda$CDM limit, we obtain $c_{1}=2.0\times10^{6}$ and $c_{2}=1.5\times10^{5}$. Considering these values for $c_{1}$ and $c_{2}$, in Fig.~\ref{fig:HS-background-potential} we show the profile of the HS scalar field potential in the Jordan frame, noting a slow evolution of $V\left(\phi\right)$. 

\section{The LTB model in General Relativity}\label{sec: LTB in GR}

The LTB spherical solution \citep{Lemaitre:1927zz,Tolman:1934za,Bondi:1947fta,Peebles:2002gy,Montani-book2009hju} describes the geometry of an inhomogeneous but isotropic Universe, generalizing the FLRW line element \citep{Weinberg:2008zzc}. In the LTB model, the space is isotropic only observing the Universe from a specific preferred point, i.e., the center that is singled out by adopting a spherical symmetry, where an observer is supposed to be located. Hereinafter, we consider the evolution of a dust cosmological model resulting in a spherical mass overdensity (or underdensity) with vanishing pressure ($p=0$), which can be formulated in the LTB formalism. Furthermore, we focus on the late Universe, hence we neglect relativistic species, since they are subdominant today, i.e., $\Omega_{r0}\sim 10^{-5}$.

The LTB spherically symmetric line element in the synchronous gauge is written as
\begin{equation}
ds^{2}=-dt^{2}+e^{2\alpha}dr^{2}+e^{2\beta}\,d\Omega^{2}\,,\label{eq:LTBmetric}
\end{equation}
in which $t$ is the cosmic time, $r$ is the radial coordinate indicating the spatial distance from the preferred point, and $d\Omega$ is the solid angle element. Note that there are two metric functions: $\alpha=\alpha\left(t,\,r\right)$ and $\beta=\beta\left(t,\,r\right)$. 

The three independent Einstein field equations in the $\Lambda$LTB model with a pressure-less dust are provided by the 0-1, 0-0, and 1-1 components, which rewrite as
\begin{subequations}
\begin{align}
 & \frac{\dot{\beta}^{\prime}}{\beta^{\prime}}-\dot{\alpha}+\dot{\beta}=0\,,\label{subeq:LTB-01-GR}\\
 & \dot{\beta}^{2}+2\dot{\alpha}\dot{\beta}+e^{-2\beta}-e^{-2\alpha}\left[2\beta^{\prime\prime}+3\beta^{\prime}{}^{2}-2\alpha^{\prime}\beta^{\prime}\right]=\nonumber\\
 & = \chi\rho+\Lambda,\label{subeq:LTB-00-GR}\\
 & 2\,\ddot{\beta}+3\,\dot{\beta}^{2}+e^{-2\beta}-\beta^{\prime}{}^{2}\,e^{-2\alpha}=\Lambda\,,\label{subeq:LTB-11-GR}
\end{align}
\end{subequations}
respectively, where $\dot{\left(\right)}=d/dt$ and $\left(\right){}^{\prime}=d/dr$. We have separated explicitly the cosmological constant and matter term, and $\rho$ is simply the energy density of the matter component (we neglected the subscript m in $\rho$ for brevity). The other non-null field equations in the LTB metric are related to the previous equations system due to the spherical symmetry. More precisely, it is straightforward to show the following relations between the Einstein tensor components in the LTB metric:
\begin{align}
G_{\,2}^{2} & =G_{\,1}^{1}+\frac{\left(G_{\,1}^{1}\right)^{\prime}}{2\,\beta^{\prime}}\,,\label{eq:relazioneG11-G22-GR}
\end{align}
and also $G_{\,3}^{3}=G_{\,2}^{2}$. 

It should be noted that the two metric functions $\alpha$ and $\beta$ can be related in GR, by exploiting the 0-1 component \eqref{subeq:LTB-01-GR} of the Einstein field equations, in order to rewrite the LTB line element \eqref{eq:LTBmetric} in a simpler form \citep{Peebles:1994xt}. Indeed, Eq.~\eqref{subeq:LTB-01-GR} admits the solution
\begin{equation}
\beta^{\prime}=g\left(r\right)\,e^{\alpha-\beta}\,,
\end{equation}
where $g\left(r\right)$ is an arbitrary function of the radial coordinate $r$. 
Then, the LTB line element \eqref{eq:LTBmetric} can be rewritten as
\begin{equation}
ds^{2}=-dt^{2}+\frac{\left[\left(a\,r\right){}^{\prime}\right]{}^{2}}{1-r^{2}\,K^{2}}dr^{2}+\left(a\,r\right){}^{2}\,d\Omega^{2}\,,\label{eq:LTBmetric-GR-simpler}
\end{equation}
in which the following parametrization has been adopted
\begin{equation}
g\left(r\right)\equiv\left(1-r^{2}\,K^{2}\right){}^{1/2}\label{eq:equation4.8}
\end{equation}
with $K=K\left(r\right)$, and $a\left(t,r\right)\equiv e^{\beta} r^{-1}$ has been defined as the generalization of the scale factor in an inhomogeneous Universe.

By using the LTB metric, the remaining Einstein field equations \eqref{subeq:LTB-00-GR} and \eqref{subeq:LTB-11-GR} become
\begin{subequations}
\begin{align}
 & 3\,\left[\dot{a}^{2}\,a\,r^{3}+a\,r^{3}\,K^{2}\right]{}^{\prime}=\left(\chi\,\rho+\Lambda\right)\,\left[\left(a\,r\right){}^{3}\right]{}^{\prime}\,,\label{subeq:LTB-00-GR-simpler}\\
 & \frac{2\,\ddot{a}}{a}+\frac{\dot{a}^{2}}{a^{2}}+\frac{K^{2}}{a^{2}}=\Lambda\,,\label{subeq:LTB-11-GR-simpler}
\end{align}
\end{subequations}
respectively.

It may be observed that the form of the LTB metric given by Eq.~\eqref{eq:LTBmetric-GR-simpler} reminds the FLRW line element. More specifically, if $a\left(t,r\right)$ and the LTB curvature function $K\left(r\right)$ do not depend on the radial coordinate $r$, the FLRW geometry is exactly recovered to describe a homogeneous and isotropic Universe. Furthermore, it is straightforward to show in that limit that Eqs.~\eqref{subeq:LTB-00-GR-simpler} and \eqref{subeq:LTB-11-GR-simpler} turn into the Friedmann equations in the FLRW metric.

Finally, the continuity equation for a pressure-less perfect fluid in the LTB metric \eqref{eq:LTBmetric} can be written as
\begin{equation}
\dot{\rho}+\left(\dot{\alpha}+2\,\dot{\beta}\right)\,\rho=0\,,\label{eq:continuity-LTB-general-metric}
\end{equation}
or equivalently
\begin{equation}
\dot{\rho}+\left(\frac{\dot{a}+r\,\dot{a}^{\prime}}{a+r\,a^{\prime}}+2\,\frac{\dot{a}}{a}\right)\,\rho=0\,,\label{eq:continuity-LTB-simpler-metric}
\end{equation}
if the LTB metric in the form given by Eq.~\eqref{eq:LTBmetric-GR-simpler} is considered. We recall that the energy conservation law results from the divergenceless law of the stress-energy tensor $\nabla_{\mu}T^{\mu\nu}=0$ for $\nu=0$.

\section{The LTB model in the Jordan frame of $f\left(R\right)$ gravity}\label{sec:LTB-JF}

We study the cosmological dynamics in the LTB metric \eqref{eq:LTBmetric} within the framework of the $f\left(R\right)$ gravity in the Jordan frame. The 0-1, 0-0, 1-1 components of the gravitational field equations \eqref{subeq: field equations Jordan frame - metric} are written as
\begin{subequations}
\label{subeqs:LTB-Jordan-frame}
\begin{align}
 & \frac{\dot{\beta}^{\prime}}{\beta^{\prime}}-\dot{\alpha}+\dot{\beta}=-\frac{1}{2\,\phi\,\beta^{\prime}}\,\left(\dot{\phi}^{\prime}-\dot{\alpha}\,\phi^{\prime}\right)\,,\label{subeq: 01 LTB Jordan frame}\\
 & \dot{\beta}^{2}+2\dot{\alpha}\,\dot{\beta}+e^{-2\beta}-e^{-2\alpha}\left[2\beta^{\prime\prime}+3\left(\beta^{\prime}\right)^{2}-2\alpha^{\prime}\beta^{\prime}\right]=\nonumber \\
 & \quad-\frac{1}{\phi}\left\{ \left(\dot{\alpha}+2\dot{\beta}\right)\dot{\phi}-e^{-2\alpha}\left[\phi^{\prime\prime}-\phi^{\prime}\left(\alpha^{\prime}-2\beta^{\prime}\right)\right]\right\} \nonumber \\
 & \quad+\frac{\chi\,\rho}{\phi}+\frac{V\left(\phi\right)}{2\,\phi}\,,\label{subeq: 00 LTB Jordan frame}\\
 & 2\,\ddot{\beta}+3\,\dot{\beta}^{2}+e^{-2\,\beta}-e^{-2\,\alpha}\,\left(\beta^{\prime}\right)^{2}=\frac{V\left(\phi\right)}{2\,\phi}\nonumber \\
 & \quad\,-\frac{1}{\phi}\,\left[\ddot{\phi}+2\,\dot{\beta}\,\dot{\phi}-2\,e^{-2\,\alpha}\,\beta^{\prime}\,\phi^{\prime}\right]\,,\label{subeq: 11 LTB Jordan frame}
\end{align}
\end{subequations}
respectively, where we have put a cosmological pressure-less dust as a source. In the Appendix~\ref{app:Verifying-spatial-isotropy}, it is shown that the other non-vanishing gravitational field equations, i.e., 2-2 and 3-3 components, depend on the previous set of equations, as it must be, basically due to the spherical symmetry in the LTB geometry. 

Moreover, the scalar field equation \eqref{eq: scalar field eq. Jordan frame} rewrites as
\begin{align}
 & \ddot{\phi}+\left(\dot{\alpha}+2\,\dot{\beta}\right)\,\dot{\phi}-e^{-2\,\alpha}\,\left[\phi^{\prime\prime}-\phi^{\prime}\,\left(\alpha^{\prime}-2\,\beta^{\prime}\right)\right]\nonumber \\
 & \quad-\frac{2}{3}\,V\left(\phi\right)+\frac{\phi}{3}\,\frac{dV}{d\phi}=\frac{\chi\,\rho}{3}\,.\label{eq: scalar LTB Jordan frame}
\end{align}
Note that in an inhomogeneous cosmology all the quantities $\phi$, $\rho$, $\alpha$, and $\beta$ depend on both $t$ and $r$.

It should be emphasized the occurrence of extra contributions in field equations \eqref{subeqs:LTB-Jordan-frame} and \eqref{eq: scalar LTB Jordan frame} with respect to the $\Lambda$LTB model, due to the coupling between the scalar field $\phi$ and the metric functions $\alpha$, $\beta$, as well as the presence of the scalar field potential. For instance, it is quite clear to recognize an extra coupling term by comparing the 0-1 field Eq.~\eqref{subeq: 01 LTB Jordan frame} with the respective Eq.~\eqref{subeq:LTB-01-GR} in GR. As a consequence, this coupling in the Jordan frame does not allow us to find a relation between $\alpha$ and $\beta$, and then rewrite the LTB metric in a simpler form, unlike the $\Lambda$LTB model in GR (\prettyref{sec: LTB in GR}). For all these reasons, the cosmological dynamics in the Jordan frame is really different from the GR scenario. 

On the opposite, the continuity equation related to the ordinary stress-energy tensor $T_{\mu\nu}$ for a dust in the LTB metric exactly exhibits the same form provided in Eq.~\eqref{eq:continuity-LTB-general-metric} both in GR and $f\left(R\right)$ gravity. In the latter theory, other additional equations are those related to the effective stress-energy tensor $T_{\mu\nu}^{\left[\phi\right]}$, defined in Eq.~\eqref{eq: def effective stress-energy tensor}, for the scalar field in the Jordan frame. More specifically, Eqs.~\eqref{eq: effective stress-energy tensor law} in the LTB metric become
\begin{subequations}
\label{subeqs:effective-tensor-LTB}
\begin{align}
 & \frac{1}{2}\,\frac{dV}{d\phi}-\frac{V\left(\phi\right)}{2\,\phi}-\ddot{\alpha}-2\,\ddot{\beta}-\dot{\alpha}^{2}-2\,\dot{\beta}^{2}=\frac{\chi}{\phi}\,\rho\nonumber \\
 & \,\,-\frac{1}{\phi}\left\{ \left(\dot{\alpha}+2\dot{\beta}\right)\dot{\phi}-e^{-2\alpha}\left[\phi^{\prime\prime}-\phi^{\prime}\left(\alpha^{\prime}-2\beta^{\prime}\right)\right]\right\} \,,\label{subeq: effect tensor mu=00003D0 LTB}\\
 & \ddot{\alpha}+\dot{\alpha}\,\left(\dot{\alpha}+2\,\dot{\beta}\right)+2\,e^{-2\,\alpha}\,\left[\beta^{\prime}\,\left(\alpha^{\prime}-\beta^{\prime}\right)-\beta^{\prime\prime}\right]\nonumber \\
 & \,\,-\frac{1}{\phi}\,\left[\ddot{\phi}\,+2\,\dot{\beta}\,\dot{\phi}-2\,e^{-2\,\alpha}\,\beta^{\prime}\,\phi^{\prime}\right]+\frac{V\left(\phi\right)}{2\,\phi}=\frac{1}{2}\,\frac{dV}{d\phi}\label{subeq: effect tensor mu=00003D1 LTB}
\end{align}
\end{subequations}
for $\mu=0,\,1$, respectively. 

It should be recalled that these laws for $T_{\mu\nu}^{\left[\phi\right]}$ are not independent of the field equations \eqref{subeqs:LTB-Jordan-frame} and \eqref{eq: scalar LTB Jordan frame}, because these laws basically come from field equations using Bianchi identities. Nevertheless, these additional equations \eqref{subeqs:effective-tensor-LTB} can be useful to rewrite field equations in a different form. For instance, Eq.~\eqref{subeq: effect tensor mu=00003D1 LTB} has been employed to find the dependence between the 1-1 and 2-2 components of the gravitational field equations \eqref{subeq: 11 LTB Jordan frame} and \eqref{eq: 22 LTB Jordan frame-app}, which implies spatial isotropy in the LTB geometry, as it has been shown in the Appendix~\ref{app:Verifying-spatial-isotropy}.

To sum up, within the Jordan frame of $f\left(R\right)$ gravity in the LTB metric, we have obtained a system of four partial differential equations \eqref{subeqs:LTB-Jordan-frame} and \eqref{eq: scalar LTB Jordan frame} with four unknown functions: $\alpha\left(t,r\right)$, $\beta\left(t,r\right)$, $\rho\left(t,r\right)$, and $\phi\left(t,r\right)$. Note that the scalar field potential $V\left(\phi\right)$ provides a degree of freedom in the theory. Furthermore, other supplementary equations are provided by Eqs.~\eqref{eq:continuity-LTB-general-metric} and \eqref{subeqs:effective-tensor-LTB}.

\section{Perturbation approach for the LTB model in General Relativity}\label{sec:pert-approach-GR}

In this section, we consider local inhomogeneities of the Universe as small spherically symmetric perturbations over a flat background FLRW geometry. We follow a linear perturbation approach, so that we have the FLRW geometry at the zeroth-order perturbation theory, while we build a lumpy Universe described by the LTB metric at the first-order perturbation \citep{Marcoccia:2018anj}. Thus, we can write the LTB metric tensor components as
\begin{equation}
g_{\mu\nu}^{\textrm{LTB}}=\bar{g}_{\mu\nu}^{\textrm{FLRW}}+\delta g_{\mu\nu}\,.\label{eq:metric-decomposition}
\end{equation}
Hereinafter, we use an overbar to denote quantities referred to the background homogeneous and isotropic Universe, and the symbol $\delta$ is related to linear perturbation terms. We emphasize that, choosing this decomposition in Eq.~\eqref{eq:metric-decomposition}, we require spherically symmetric perturbations.

Moreover, we adopt the synchronous gauge for the LTB metric, which intrinsically includes two degrees of freedom in the perturbed metric. Indeed, two independent metric functions, i.e., $\alpha\left(t,r\right)$ and $\beta\left(t,r\right)$, are contained in the original LTB line element~\eqref{eq:LTBmetric} or, equivalently, $a\left(t,r\right)$ and $K^{2}\left(r\right)$ in GR, according to Eq.~\eqref{eq:LTBmetric-GR-simpler}. Actually, $K^{2}\left(r\right)$ is not exactly a dynamical degree of freedom but an arbitrary parametric function, as a result of the LTB cosmological dynamics. To be more specific, we recall that, in \prettyref{sec: LTB in GR}, the 0-1 component \eqref{subeq:LTB-01-GR} of the Einstein field equations has allowed us to find a relation between the metric functions $\alpha\left(t,r\right)$ and $\beta\left(t,r\right)$ in GR, hence to reduce one degree of freedom, and the LTB metric in GR assumes the form given in Eq.~\eqref{eq:LTBmetric-GR-simpler} in terms of $a\left(t,r\right)$ and $K^{2}\left(r\right)$.

Since the background FLRW and the perturbed LTB metrics are both locally rotationally symmetric and are given in the same normal geodesic frame, we only need to focus on scalar functions, as shown in \citep{vanElst:1995eg,Sussman:2010jf}. Hence, the scale factor $a\left(t,r\right)$ and the energy density of the matter component $\rho\left(t,r\right)$, contained in the field equations~\eqref{subeq:LTB-00-GR-simpler} and \eqref{subeq:LTB-11-GR-simpler}, are defined as:
\begin{subequations}
\begin{align}
a\left(t,r\right) & =\bar{a}\left(t\right)+\delta a\left(t,r\right)\,,\label{subeq:decomposition-pert-scale-factor-GR}\\
\rho\left(t,r\right) & =\bar{\rho}\left(t\right)+\delta\rho\left(t,r\right)\,.\label{subeq:decomposition-pert-density-GR}
\end{align}
\end{subequations}

Note that the curvature function $K^{2}\left(r\right)$ in the LTB metric \eqref{eq:LTBmetric-GR-simpler} involves radial inhomogeneities, so it can be regarded as a perturbative contribution. In other words, even if hereinafter we set the curvature parameter in the FLRW metric as $\bar{k}=0$, it is possible to define a linear curvature pertubation, which is exactly given by the curvature function $K^{2}\left(r\right)$ in the LTB metric, according to the metric decomposition in Eq.~\eqref{eq:metric-decomposition}. Thus, in addition to the metric perturbation $\delta a$, we have another perturbed quantity with respect to the background FLRW metric, which is $K^{2}\left(r\right)$. 

We also require that background terms dominate over the linear perturbations
\begin{equation}
\delta a\left(t,r\right)\ll\bar{a}\left(t\right)\,,\quad\qquad\delta\rho\left(t,r\right)\ll\bar{\rho}\left(t\right)
\end{equation}
at any time $t$, or redshift $z$, in the late Universe. This condition is dictated by the cosmological principle. 

Once the decomposition has been defined in Eqs.~\eqref{subeq:decomposition-pert-scale-factor-GR}, \eqref{subeq:decomposition-pert-density-GR}, the evolution in time and space of the physical quantities can be obtained by studying the gravitational field equations~\eqref{subeq:LTB-00-GR-simpler} and \eqref{subeq:LTB-11-GR-simpler} at background and linear levels. 

In particular, we define a dimensionless time variable
\begin{equation}
\tau=\frac{t}{t_{0}}\,,\label{eq:definition_tau}
\end{equation}
in which $t_{0}$ is the present cosmic time (today $\tau=1$). Note that $\tau$ is the cosmic time in units of the present Hubble time since $t_{0}$ can be approximately written in terms of the Hubble constant as $t_{0}\approx1/H_{0}$ \citep{Weinberg:2008zzc}.

\subsection{Background solution}\label{subsec:Background-solution}

If we consider only background terms, it is straightforward to show that Eq.~\eqref{subeq:LTB-00-GR-simpler} turns into the first Friedmann equation in a flat FLRW geometry:
\begin{equation}
H^{2}\left(t\right)\equiv\left[\frac{\dot{\overline{a}}\left(t\right)}{\overline{a}\left(t\right)}\right]^{2}=\frac{\chi\,\overline{\rho}\left(t\right)}{3}+\frac{\Lambda}{3}\,,\label{eq:standard-Friedmann-eq-t-LCDM}
\end{equation}
being $H(t)$ the Hubble parameter.
Furthermore, the other Eq.~\eqref{subeq:LTB-11-GR-simpler} can be rewritten at the background level, and combined with Eq.~\eqref{eq:standard-Friedmann-eq-t-LCDM}, provides the second Friedmann equation, also commonly named the cosmic acceleration equation:
\begin{equation}
\frac{\ddot{\overline{a}}\left(t\right)}{\overline{a}\left(t\right)}=-\frac{\chi\,\overline{\rho}}{6}+\frac{\Lambda}{3}\,.\label{eq:standard-acc-eq-t-LCDM}
\end{equation}
Then, we rewrite the first Friedmann equation~\eqref{eq:standard-Friedmann-eq-t-LCDM} in terms of $\tau$, defined in Eq.~\eqref{eq:definition_tau}, as
\begin{equation}
\left[\frac{1}{\bar{a}\left(\tau\right)}\,\frac{d\bar{a}\left(\tau\right)}{d\tau}\right]^{2}=\frac{\Omega_{m0}}{\bar{a}^{3}\left(\tau\right)}+\Omega_{\Lambda0}\,.\label{eq:Friedmann-eq-tau-LCDM}
\end{equation}
We have used the well-known relation $\bar{\rho}\sim \bar{a}^{-3}$ for the matter component, the chain rule $\frac{d}{dt}=\frac{1}{t_{0}}\,\frac{d}{d\tau}\approx H_{0}\,\frac{d}{d\tau}$, and the usual definitions of the cosmological density parameters $\Omega_{m0}$ and $\Omega_{\Lambda0}$.

The Friedmann equation~\eqref{eq:Friedmann-eq-tau-LCDM} admits an analytical solution in the late Universe, that is the background scale factor in terms of $\tau$:
\begin{align}
 & \bar{a}\left(\tau\right) =\left(\frac{\Omega_{m0}}{\Omega_{\Lambda0}}\right)^{1/3}\left\{ \sinh\left[\frac{3}{2}\sqrt{\Omega_{\Lambda0}}\left(\tau-1\right)\right.\right.\nonumber\\
 & \left.\left.+\textrm{arcsinh}\left(\sqrt{\frac{\Omega_{\Lambda0}}{\Omega_{m0}}}\right)\right]\right\} ^{2/3}.\label{eq:background-scale-factor-GR} 
\end{align}
Note also from Eq.~\eqref{eq:background-scale-factor-GR} that the deceleration parameter $\bar{q}\left(\tau\right)\equiv-\ddot{\bar{a}}\,\bar{a}^{-1}\,H^{-2}\rightarrow-1$ for $\tau\rightarrow+\infty$, as it should be in a dark-energy-fully-dominated Universe. We here stress that the solution~\eqref{eq:background-scale-factor-GR} applies only in the late Universe, otherwise we need to solve numerically the field equations, if we also consider the radiation contribution. Nevertheless, we are interested in the evolution of local inhomogeneities at late times.

It should be noted that in the limit
\begin{equation}
\underset{\Omega_{\Lambda}\rightarrow0}{lim}\quad\bar{a}\left(\tau\right)\sim\left(\tau-1\right)^{2/3},\label{eq:equation5.7}
\end{equation}
we recover the well-known relation for the background scale factor $\bar{a}$ satisfied in the matter-dominated Universe \citep{Weinberg:2008zzc}.

Concerning the evolution of the background energy density $\bar{\rho}$, we start from the continuity equation~\eqref{eq:continuity-LTB-simpler-metric} in the LTB metric. We verify that this equation at the zeroth order becomes simply the respective continuity equation for the matter component in the FLRW metric, \textit{i.e.}
\begin{equation}
\dot{\bar{\rho}}+3\,H\,\bar{\rho}=0\,,\label{eq:continuity-background-LCDM}
\end{equation}
which provides $\bar{\rho}\sim \bar{a}^{-3}$. Actually, we recall that Eq.~\eqref{eq:continuity-background-LCDM} is not independent of the two Friedmann equations \eqref{eq:standard-Friedmann-eq-t-LCDM} and \eqref{eq:standard-acc-eq-t-LCDM}, since it can be derived by combining them.

\begin{figure}
\centering \includegraphics[scale=0.24]{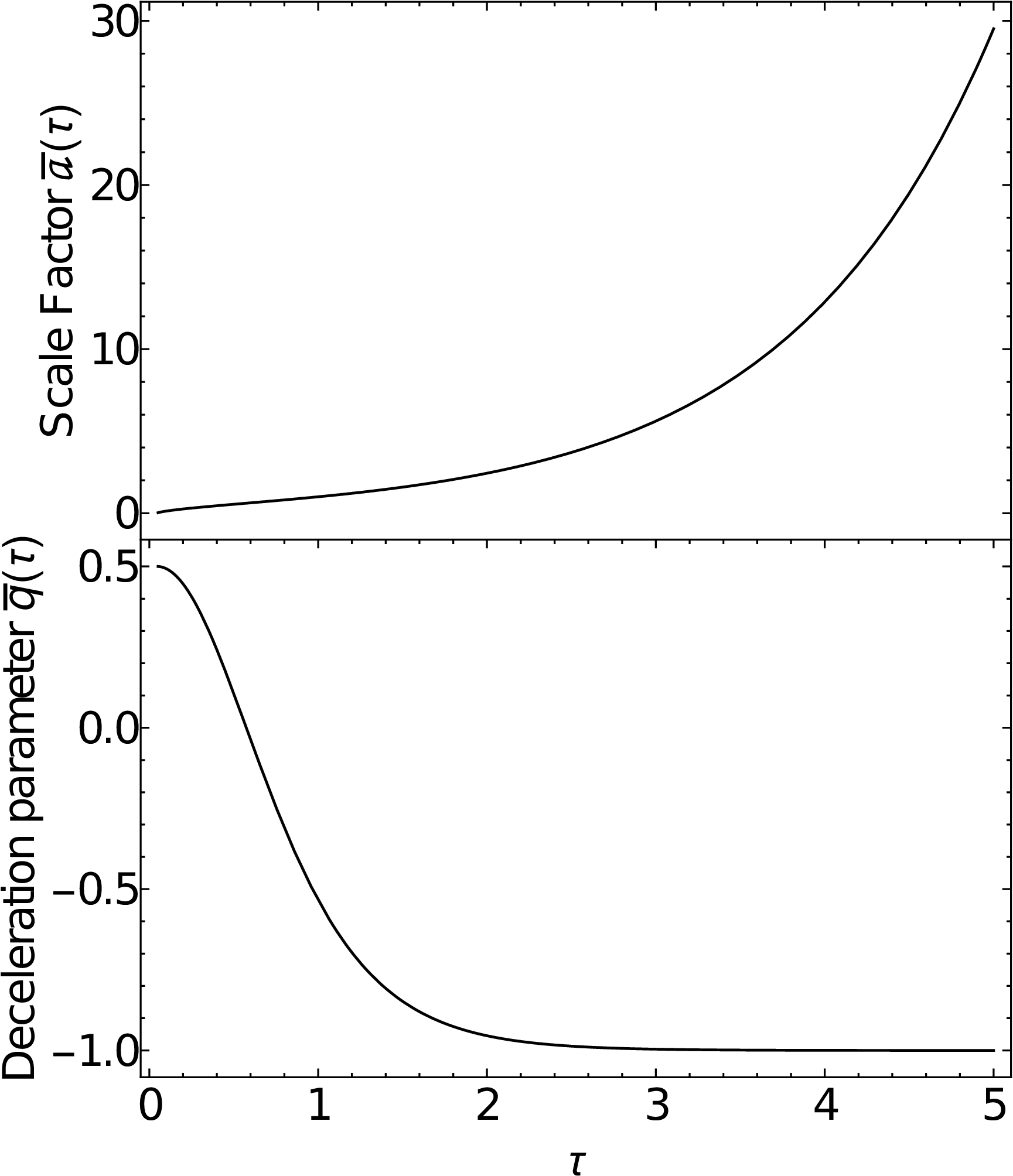}

\caption{Evolution of the background scale factor $\bar{a}\left(\tau\right)$ (top panel) and the respective deceleration parameter $\bar{q}\left(\tau\right)$ (bottom panel) in terms of the time dimensionless parameter $\tau$, defined in Eq.~\eqref{eq:definition_tau}, for a flat $\Lambda$CDM model within the range $\tau_{100\textrm{eq}}<\tau<5$, according to the solution given in Eq.~\eqref{eq:background-scale-factor-GR}.}
\label{fig:background-scale-factor-and-q}
\end{figure}

Finally, we focus on the relation between the redshift $z$ and the dimensionless parameter $\tau$, defined in Eq.~\eqref{eq:definition_tau}, to understand to what extent of $\tau$ values the background solution~\eqref{eq:background-scale-factor-GR} can be applied in the late Universe. To be more accurate in this computation, we also include the radiation contribution to write $\tau$ as
\begin{equation}
\tau\left(z\right)=1+\int_{1}^{\nicefrac{1}{1+z}}\frac{dx}{x\sqrt{\Omega_{m0}x^{-3}+\Omega_{r0}x^{-4}+\Omega_{\Lambda0}}}\,,\label{eq:tau(z)}
\end{equation}
where we have used the Friedmann equation~\eqref{eq:Friedmann-eq-tau-LCDM}, the definition of $\tau$ in Eq.~\eqref{eq:definition_tau}, and we recall that $\bar{a}_{0}/\bar{a}=1+z$. The quantity $\tau\left(z\right)$ can be computed numerically for a given redshift $z$, after specifying the cosmological parameters: $\Omega_{m0}=0.3111$, $\Omega_{\Lambda0}=0.6889$, and $\Omega_{r0}=9.138\times10^{-5}$ from Table 2 in \citep{Planck:2018vyg}. For instance, we can compute: $\tau\left(z_{\textrm{eq}}\right)=0.046$ at the redshift of the matter-radiation equality $z_{\textrm{eq}}=3403.5$; $\tau\left(z_{\textrm{DE}}\right)=0.75$ at the matter-dark energy equality $z_{\textrm{DE}}=0.303$; $\tau\left(z_{\textrm{100eq}}\right)=0.052$ when the energy density of the matter component was one hundred times more than the radiation contribution at $z_{100\textrm{eq}}=33.045$. In particular, considering this latter value of $\tau$, you can see in Fig.~\ref{fig:background-scale-factor-and-q} the behavior of the background scale factor $\bar{a}\left(\tau\right)$ and the deceleration parameter $\bar{q}\left(\tau\right)$ in the range $\tau_{100\textrm{eq}}<\tau<5$, when relativistic species are negligible.

\subsection{Linearly perturbed solutions in an inhomogeneous Universe}

We analyze the impact of local inhomogeneities in the cosmological dynamics. The linearized field equations allow us to investigate the evolution of spherically symmetric perturbations. 

If we include local inhomogeneities in the first-order perturbation theory, the Eqs.~\eqref{subeq:LTB-00-GR-simpler} and \eqref{subeq:LTB-11-GR-simpler} become
\begin{subequations}
\begin{align}
 & \left[\bar{a}^{2}\,\left(\chi\,\bar{\rho}+\Lambda\right)-\dot{\bar{a}}^{2}\right]\,\left(3\,\delta a+r\,\delta a^{\prime}\right)+\chi\,\delta\rho\,\bar{a}^{3}=\nonumber \\
 & \qquad2\,\bar{a}\,\dot{\bar{a}}\,\left(3\,\delta\dot{a}+r\,\delta\dot{a}^{\prime}\right)+3\,\bar{a}\,K^{2}+2\,\bar{a}\,r\,K\,K^{\prime}\,,\label{subeq:00-GR-perturb}\\
 & \delta\ddot{a}+\frac{\dot{\bar{a}}}{\bar{a}}\,\delta\dot{a}-\left(\frac{\ddot{\bar{a}}}{\bar{a}}+\frac{\dot{\bar{a}}^{2}}{\bar{a}^{2}}\right)\,\delta a+\frac{K^{2}}{2\,\bar{a}}=0\,,\label{subeq:11-GR-perturb}
\end{align}
\end{subequations}
respectively. 

\noindent Furthermore, we rewrite the continuity equation~\eqref{eq:continuity-LTB-simpler-metric} at linear order:
\begin{equation}
\delta\dot{\rho}+3\frac{\dot{\bar{a}}}{\bar{a}}\delta\rho=\frac{\bar{\rho}}{\bar{a}^{2}}\left[\dot{\bar{a}}\left(3\delta a+r\delta a^{\prime}\right)-\bar{a}\left(3\delta\dot{a}+r\delta\dot{a}^{\prime}\right)\right]\,.\label{eq:contin-GR-perturb}
\end{equation}

To study separately the evolution of local inhomogeneities in time and space, we adopt the separation of variables method to solve analytically the first-order perturbation equations. Hence, we define time and radial functions for all linear perturbations:
\begin{equation}
\delta a\left(t,r\right)\equiv\text{\textcyr{\cyra}}_{p}\left(t\right)\mathcal{\mathfrak{a}}_{p}\left(r\right),\,\,\,\,\,\,\,\delta\rho\left(t,r\right)\equiv R_{p}\left(t\right)\varrho_{p}\left(r\right).\label{eq:factorization-GR}
\end{equation}
The quantities $\text{\textcyr{\cyra}}_{p}\left(t\right)$ and $\mathcal{\mathfrak{a}}_{p}\left(r\right)$ are both dimensionless. We assume, without loss of generality, that $R_{p}\left(t\right)$ has the physical dimensions of an energy density as $\bar{\rho}$ and $\delta\rho$, while we treat $\varrho_{p}\left(r\right)$ like a dimensionless quantity. We also recall that the curvature perturbation $K^{2}\left(r\right)$ in the LTB metric depends only on the radial coordinate. 

We stress that if we would also include non-linear terms, then the separation of variables could not lead to a general solution. However, the linearization procedure adopted for the dynamics allows us to use a separation of variables characterized by the factorization~\eqref{eq:factorization-GR} of the time and space dependences in the linear perturbation theory. 

Using the factorization \eqref{eq:factorization-GR}, Eq.~\eqref{subeq:11-GR-perturb} can be split into two parts. By setting the radial dependence as
\begin{equation}
K^{2}\left(r\right)=\mathcal{\mathfrak{a}}_{p}\left(r\right)\,,\label{eq:11-GR-perturb-radial-component}
\end{equation}
we obtain an ordinary differential equation for the time evolution:
\begin{equation}
\ddot{\text{\textcyr{\cyra}}}_{p}+\frac{\dot{\bar{a}}}{\bar{a}}\,\dot{\text{\textcyr{\cyra}}}_{p}-\left[\frac{\ddot{\bar{a}}}{\bar{a}}+\left(\frac{\dot{\bar{a}}}{\bar{a}}\right)^{2}\right]\,\text{\textcyr{\cyra}}_{p}=0\,,\label{eq:11-GR-perturb-time-component}
\end{equation}
in which we have also considered that $\bar{a}\gg\text{\textcyr{\cyra}}_{p}$. We would emphasize that the assumption given in Eq.~\eqref{eq:11-GR-perturb-radial-component} is suggested by the form of Eq.~\eqref{subeq:11-GR-perturb}, once we used the separation of variables from Eq.~\eqref{eq:factorization-GR}. Nevertheless, we have still two metric perturbations given by the quantities $\text{\textcyr{\cyra}}_{p}\left(t\right)$ and $\mathcal{\mathfrak{a}}_{p}\left(r\right)$.

We proceed similarly for the first-order perturbation continuity equation~\eqref{eq:contin-GR-perturb}. After straightforward calculations, by using again the factorization~\eqref{eq:factorization-GR} in Eq.~\eqref{eq:contin-GR-perturb}, we separate terms that depend only on $t$ from those related to $r$. In particular, the time evolution is provided by
\begin{equation}
\dot{R}_{p}+3\,\frac{\dot{\bar{a}}}{\bar{a}}\,R_{p}=X\,\frac{\bar{\rho}}{\bar{a}}\,\left(\frac{\dot{\bar{a}}}{\bar{a}}\,\text{\textcyr{\cyra}}_{p}-\dot{\text{\textcyr{\cyra}}}_{p}\right)\,,\label{eq:contin-pert-GR-tau-behavior}
\end{equation}
while we obtain the following radial dependence
\begin{equation}
\varrho_{p}=\frac{1}{X}\,\left(3\,\mathcal{\mathfrak{a}}_{p}+r\,\mathcal{\mathfrak{a}}_{p}^{\prime}\right)\,.\label{eq:contin-pert-GR-r-behavior}
\end{equation}
The constant $X$ is introduced by the separation of variables method. We recall that the background solution for $\bar{a}$ is written in Eq.~\eqref{eq:background-scale-factor-GR} and also $\bar{\rho}\sim \bar{a}^{-3}$. 

We rewrite the first-order perturbation Eq.~\eqref{subeq:00-GR-perturb}, by employing Eq.~\eqref{eq:factorization-GR} and dividing both sides of the equation by $\bar{a}^{2}\,\text{\textcyr{\cyra}}_{p}\,\mathcal{\mathfrak{a}}_{p}$, as
\begin{align}
 & \left[\chi\,\bar{\rho}+\Lambda-\left(\frac{\dot{\bar{a}}}{\bar{a}}\right)^{2}-2\,\frac{\dot{\bar{a}}}{\bar{a}}\,\frac{\dot{\text{\textcyr{\cyra}}}_{p}}{\text{\textcyr{\cyra}}_{p}}\right]\,\left(3+r\,\frac{\mathcal{\mathfrak{a}}_{p}^{\prime}}{\mathcal{\mathfrak{a}}_{p}}\right)=\nonumber \\
 & \qquad\,\frac{3}{\bar{a}\,\text{\textcyr{\cyra}}_{p}}\,\frac{K^{2}}{\mathcal{\mathfrak{a}}_{p}}+\frac{2}{\bar{a}\,\text{\textcyr{\cyra}}_{p}}\,\frac{r\,K\,K^{\prime}}{\mathcal{\mathfrak{a}}_{p}}-\chi\,\bar{a}\,\frac{R_{p}}{\text{\textcyr{\cyra}}_{p}}\,\frac{\varrho_{p}}{\mathcal{\mathfrak{a}}_{p}}\,.\label{eq:00-interm}
\end{align}
We observe the presence of several mixed terms depending both on $t$ and $r$. However, we can reduce the number of these mixed terms, by using Eqs.~\eqref{eq:standard-Friedmann-eq-t-LCDM}, \eqref{eq:11-GR-perturb-radial-component}, and \eqref{eq:contin-pert-GR-r-behavior} to rewrite Eq.~\eqref{eq:00-interm} as
\begin{equation}
\left[2\,\frac{\dot{\bar{a}}}{\bar{a}}\,\left(\frac{\dot{\bar{a}}}{\bar{a}}\,\text{\textcyr{\cyra}}_{p}-\dot{\text{\textcyr{\cyra}}}_{p}\right)+\frac{\chi}{X}\,\bar{a}\,R_{p}\right]\,\frac{\varrho_{p}}{\mathcal{\mathfrak{a}}_{p}}=0\,.\label{eq:00-interm2}
\end{equation}

\noindent Since we want to avoid trivial solutions, we should have $\varrho_{p}\neq0$ and $\mathcal{\mathfrak{a}}_{p}\neq0$. Then, we obtain a single ordinary differential equation in the time domain:
\begin{equation}
2\,\frac{\dot{\bar{a}}}{\bar{a}}\,\left(\frac{\dot{\bar{a}}}{\bar{a}}\,\text{\textcyr{\cyra}}_{p}-\dot{\text{\textcyr{\cyra}}}_{p}\right)+\frac{\chi}{X}\,\bar{a}\,R_{p}=0\,.\label{eq:00-pert-GR-t-behavior}
\end{equation}

It is straightforward to check the compatibility between Eqs.~\eqref{eq:11-GR-perturb-time-component}, \eqref{eq:contin-pert-GR-tau-behavior}, and \eqref{eq:00-pert-GR-t-behavior}. Indeed, by combining the time derivative of Eq.~\eqref{eq:00-pert-GR-t-behavior} with Eq.~\eqref{eq:contin-pert-GR-tau-behavior}, the background field equations~\eqref{eq:standard-Friedmann-eq-t-LCDM}, and \eqref{eq:standard-acc-eq-t-LCDM}, it is easy to build exactly Eq.~\eqref{eq:11-GR-perturb-time-component}. 

Then, we rewrite the term in the brackets on the right-hand side of Eq.~\eqref{eq:contin-pert-GR-tau-behavior} by using Eq.~\eqref{eq:00-pert-GR-t-behavior}, and we get an ordinary differential equation with a single variable $R_{p}\left(t\right)$, that is
\begin{equation}
\dot{R}_{p}+\left(3\,\frac{\dot{\bar{a}}}{\bar{a}}+\frac{\chi\,\bar{\rho}}{2}\,\frac{\bar{a}}{\dot{\bar{a}}}\right)\,R_{p}=0\,.\label{eq:GR-pert-Rp}
\end{equation}

\begin{figure}[ht]
\centering \includegraphics[scale=0.24]{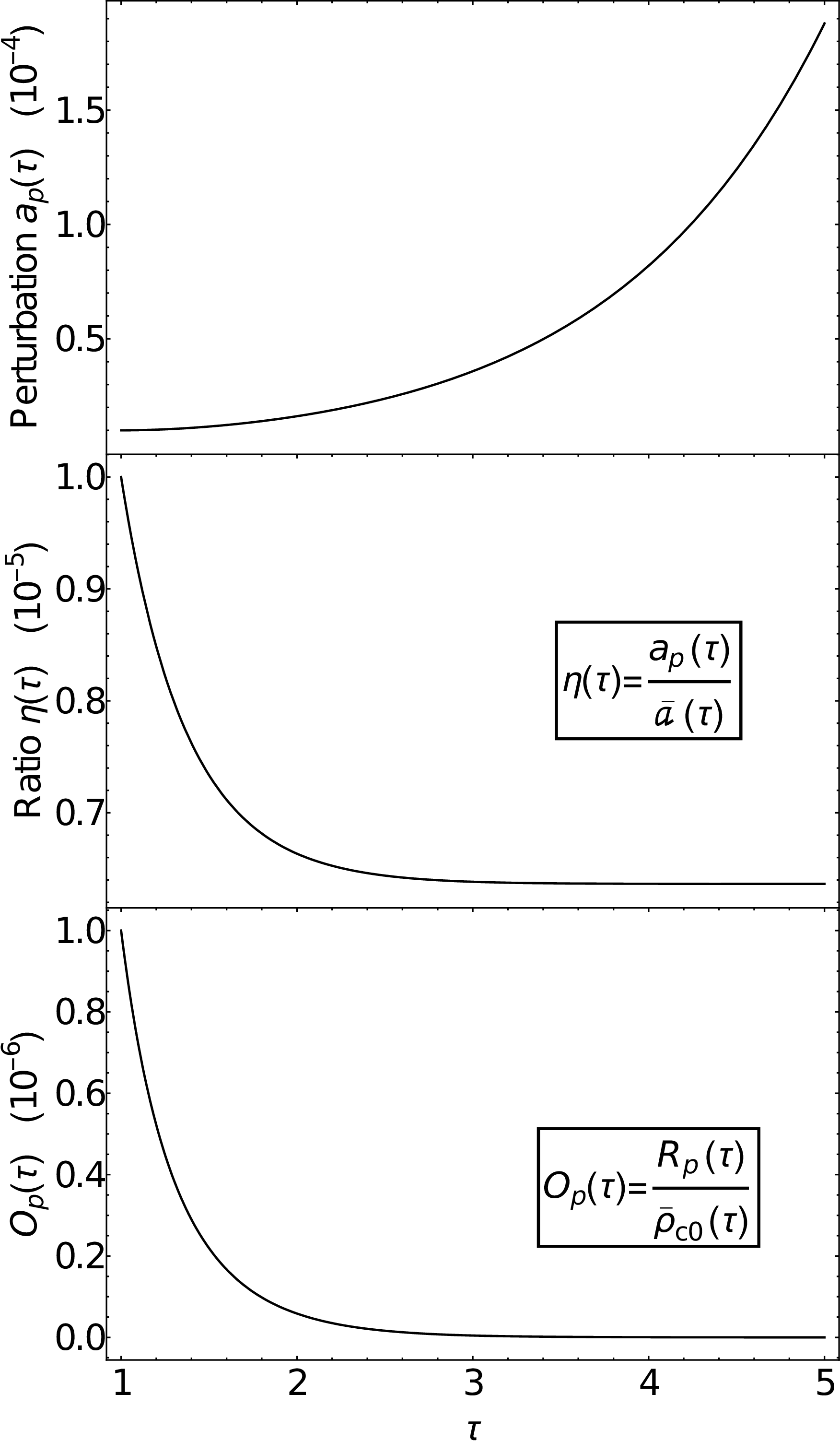}

\caption{Top panel: evolution of the linearly perturbed scale factor $\text{\textcyr{\cyra}}_{p}\left(\tau\right)$ in units of $10^{-4}$ and in terms of the parameter $\tau$, provided by the numerical solution of Eq.~\eqref{eq:11-GR-perturb-time-component}. Middle panel: the ratio between the first-order perturbation term and background scale factor $\eta\left(\tau\right)\equiv\left|\text{\textcyr{\cyra}}_{p}/\bar{a}\right|$ versus $\tau$ in units of $10^{-5}$. Bottom panel: evolution of the dimensionless perturbed energy density $O_{p}\left(\tau\right)\equiv R_{p}/\bar{\rho}_{c0}$ in units of $10^{-6}$. Note that all perturbed contributions are smaller than respective background terms for $1\protect\leq\tau\protect\leq5$.}
\label{fig:pertGR}
\end{figure}

Therefore, focusing on the time domain, we can obtain numerical solutions for the unknown quantities $\text{\textcyr{\cyra}}_{p}$ and $R_{p}$ from the linearized equations~\eqref{eq:11-GR-perturb-time-component} and \eqref{eq:GR-pert-Rp}. In particular, Eq.~\eqref{eq:11-GR-perturb-time-component} shows exactly the same behavior in terms of $t$ and $\tau$, as it can be checked by using the definition of $\tau$~\eqref{eq:definition_tau}. Recalling the expression~\eqref{eq:background-scale-factor-GR} of the background scale factor in GR, Eq.~\eqref{eq:11-GR-perturb-time-component} can be solved numerically. We set the initial conditions at $\tau=1$ today: $\text{\textcyr{\cyra}}_{p}\left(1\right)=10^{-5}$ and $\dot{\text{\textcyr{\cyra}}}_{p}\left(1\right)=0$. Moreover, we fixed the same values for $\Omega_{m0}$ and $\Omega_{\Lambda0}$ adopted in \prettyref{subsec:Background-solution}. In the upper panel of Fig.~\ref{fig:pertGR}, you can see the numerical results for $1\leq\tau\leq5$. Note that the perturbed scale factor $\text{\textcyr{\cyra}}_{p}$ increases as $\tau$ grows, and this fact may be a problem if perturbations become unstable. However, the evolution of $\text{\textcyr{\cyra}}_{p}$ is dominated by the background term $\bar{a}$ at any time $\tau$. Indeed, the ratio between the perturbation and background terms with $\eta\left(\tau\right)\equiv\left|\text{\textcyr{\cyra}}_{p}/\bar{a}\right|\ll1$ for any $\tau$, as it is shown in the middle panel of Fig.~\ref{fig:pertGR}. In other words, perturbations of the scale factor due to local inhomogeneities will remain small over time. 

Concerning the time evolution of the perturbed energy density of the matter component $R_{p}$, we rewrite Eq.~\eqref{eq:GR-pert-Rp} in terms of $\tau$ as
\begin{equation}
\frac{dO_{p}}{d\tau}+3\,\left[\frac{1}{\bar{a}}\,\frac{d\bar{a}}{d\tau}+\frac{\Omega_{m0}}{2\,\bar{a}^{2}}\,\left(\frac{d\bar{a}}{d\tau}\right)^{-1}\right]\,O_{p}=0\,,\label{eq:GR-pert-Op}
\end{equation}
in which we have defined the dimensionless quantity $O_{p}\equiv R_{p}/\bar{\rho}_{c0}$. We solve numerically Eq.~\eqref{eq:GR-pert-Op} with the initial condition $O_{p}\left(1\right)=10^{-5}$ for $\tau=1$. The results are shown in the bottom panel of Fig.~\ref{fig:pertGR}. Note that $O_{p}\left(\tau\right)\ll1$ for $1\leq\tau\leq5$. 

Finally, after obtaining numerical solutions for the time domain, we focus on the radial part of perturbations. We recall that $K^{2}\left(r\right)=\mathcal{\mathfrak{a}}_{p}\left(r\right)$ from Eq.~\eqref{eq:11-GR-perturb-radial-component}. Moreover, we assume for simplicity a proportionality between radial perturbations, i.e., $\varrho_{p}=C\,\mathcal{\mathfrak{a}}_{p}$ with a constant $C$. 

Then, we solve the ordinary differential equation~\eqref{eq:contin-pert-GR-r-behavior} in $r$, and we obtain a power-law behavior for the radial correction of the perturbed scale factor in GR:
\begin{equation}
\mathcal{\mathfrak{a}}_{p}\left(r\right)\propto r^{-y}\,,\label{eq:soluz-pert-scale-factor-GR}
\end{equation}
for which $y\equiv3-C\,X$. More in detail, we need to impose the condition $y>0$ to ensure that inhomogeneities decay on large scales, according to the cosmological principle. 

\section{Perturbation approach for the LTB model in the Jordan Frame of $f\left(R\right)$ gravity}\label{sec:Perturbation-approach-JF}

In this section, we compare and discuss the evolution of inhomogeneous perturbations in GR and in the Jordan frame of $f\left(R\right)$ gravity, considering again a cosmological dust ($p=0$) in the LTB geometry. A complete general solution in linear perturbation theory within the metric $f\left(R\right)$ gravity was developed in \citep{Song:2006ej,Hu:2007nk,Hu:2007pj}. Instead of proceeding with a fourth-order cosmological dynamics, here we work in the equivalent Jordan frame $f\left(R\right)$ gravity. Furthermore, as a particular case, we focus on spherically symmetric perturbations, following the same perturbation approach developed in \prettyref{sec:pert-approach-GR} by using the metric decomposition in Eq.~\eqref{eq:metric-decomposition}. Hence, we split the metric functions $\alpha$ and $\beta$, the energy density $\rho$, and the scalar field $\phi$ into background terms plus linear corrections as
\begin{align}
\alpha\left(t,r\right) & =\bar{\alpha}\left(t\right)+\delta\alpha\left(t,r\right)\nonumber\\
\beta\left(t,r\right) & =\bar{\beta}\left(t,r\right)+\delta\beta\left(t,r\right)\nonumber \\
\rho\left(t,r\right) & =\bar{\rho}\left(t\right)+\delta\rho\left(t,r\right)\nonumber\\
\phi\left(t,r\right) & =\bar{\phi}\left(t\right)+\delta\phi\left(t,r\right)\,.\label{eq:expansion-background+pert-Jordan}
\end{align}
We also require again that inhomogeneities are much smaller than respective background terms. 

It should be stressed that, in the Jordan frame, we can not use the LTB metric in the simpler form \eqref{eq:LTBmetric-GR-simpler}, but we refer to the original LTB line element \eqref{eq:LTBmetric}. Thus, the two degrees of freedom of the perturbed metric are given by $\alpha\left(t,r\right)$ and $\beta\left(t,r\right)$; we no longer refer to $a\left(t,r\right)$ and $K^{2}\left(r\right)$, as in the GR scenario.

Note that the background quantities $\bar{\alpha}$ and $\bar{\beta}$ are related to the scale factor $\bar{a}$, since we want to reproduce a flat FLRW geometry at the zeroth-order perturbation. Then, comparing a flat FLRW metric with the LTB line element in the form given by Eq.~\eqref{eq:LTBmetric}, it is straightforward to show that
\begin{equation}
\bar{\alpha}\left(t\right)=\ln\left(\bar{a}\left(t\right)\right)\,,\quad\quad\bar{\beta}\left(t,r\right)=\ln\left(\bar{a}\left(t\right)\,r\right)\,.\label{eq:alphabar(t)-betabar(t)-a(t)}
\end{equation}

As a consequence, the metric tensor component $g_{rr}$ in the LTB metric can be approximated for $\delta\alpha\ll1/2$ as
\begin{equation}
g_{rr}=e^{2\alpha}=e^{2\left(\bar{\alpha}+\delta\alpha\right)}\approx\bar{a}^{2}\left(t\right)\,\left(1+2\,\delta\alpha\right)\,,
\end{equation}
in which it is possible to recognize the respective metric tensor component $g_{rr}$ in a flat FLRW metric as background term. We can follow the same reasoning for the other metric tensor components involving $\beta$ with the assumption $\delta\beta\ll1/2$.

Also, we need to expand the scalar field potential $V\left(\phi\right)$, defined in Eq.~\eqref{eq: def scalar field potential}, which appears in the field equations~\eqref{subeq: 00 LTB Jordan frame}, \eqref{subeq: 11 LTB Jordan frame}, \eqref{eq: scalar LTB Jordan frame}. Therefore, including local inhomogeneities, we obtain
\begin{align}
 & V\left[\phi\left(t,r\right)\right] =V\left[\bar{\phi}\left(t\right)+\delta\phi\left(t,r\right)\right]\approx\nonumber \\
 &\quad \approx V\left[\bar{\phi}\left(t\right)\right]+\left.\frac{dV}{d\phi}\right|_{\phi=\bar{\phi}}\,\delta\phi\left(t,r\right)+O\left(\delta\phi^{2}\right)\label{eq:expand-potential}
\end{align}
for the first-order perturbation theory. Similarly, we can rewrite the derivative of $V\left(\phi\right)$.

As we have proceeded to study the gravitational field equations in GR in \prettyref{sec:pert-approach-GR} to find background and linear solutions, now we analyze the dynamics of $f\left(R\right)$ gravity in the Jordan frame, provided by the field equations \eqref{subeqs:LTB-Jordan-frame}, and \eqref{eq: scalar LTB Jordan frame} in the LTB metric. 

\subsection{Background solution}\label{subsec:Background-solution-JF}

If we consider all background quantities, we do not include spherically symmetric perturbations and Eqs.~\eqref{subeq: 00 LTB Jordan frame}, \eqref{subeq: 11 LTB Jordan frame}, and \eqref{eq: scalar LTB Jordan frame} turn into field equations in the Jordan frame of $f\left(R\right)$ gravity in a flat FLRW geometry:
\begin{subequations}
\label{system-background-JF}
\begin{align}
 & H^{2}=\frac{\chi\bar{\,\rho}}{3\,\bar{\phi}}-H\,\frac{\dot{\bar{\phi}}}{\bar{\phi}}+\frac{V\left(\bar{\phi}\right)}{6\,\bar{\phi}}\,,\label{subeq:generalized-Friedmann-FLRW}\\
 & \frac{\ddot{\bar{a}}}{\bar{a}}=-\frac{\chi\,\bar{\rho}}{6\,\bar{\phi}}-\frac{H}{2}\,\frac{\dot{\bar{\phi}}}{\bar{\phi}}-\frac{1}{2}\,\frac{\ddot{\bar{\phi}}}{\bar{\phi}}+\frac{V\left(\bar{\phi}\right)}{6\,\bar{\phi}}\,,\label{subeq:generalized-acc-FLRW}\\
 & 3\ddot{\bar{\phi}}-2\,V\left(\bar{\phi}\right)+\bar{\phi}\,\frac{dV}{d\bar{\phi}}+9\,H\,\dot{\bar{\phi}}=\chi\,\bar{\rho}\,.\label{subeq:scalar-field-FLRW}
\end{align}
\end{subequations}
Eq.~\eqref{subeq:generalized-Friedmann-FLRW} is the modified Friedmann equation, Eq.~\eqref{subeq:generalized-acc-FLRW} is the modified acceleration equation, and Eq.~\eqref{subeq:scalar-field-FLRW} concerns the scalar field evolution in the Jordan frame. In particular, Eq.~\eqref{subeq:generalized-acc-FLRW} is obtained by combining the zeroth-order perturbation Eqs.~\eqref{subeq: 00 LTB Jordan frame} and \eqref{subeq: 11 LTB Jordan frame}. Furthermore, Eq.~\eqref{subeq: 01 LTB Jordan frame} vanishes at the background level in the synchronous gauge, and it becomes a trivial identity. 

Then, we focus on the $f\left(R\right)$ HS model in the Jordan frame, which has been introduced in \prettyref{subsec:The-Hu-Sawicki-model}. Recalling the form of the scalar field potential $V\left(\bar{\phi}\right)$ in Eq.~\eqref{eq:scalar-field-potential-HS}, we observe that the field equations \eqref{system-background-JF} do not admit any analytical solutions, and we have to solve it numerically. 

In this regard, we rewrite the full set of equations \eqref{system-background-JF} in terms of the dimensionless parameter $\tau$, defined in Eq.~\eqref{eq:definition_tau}, and we obtain, respectively:
\begin{subequations}
\label{system-background-JF-tau}
\begin{align}
 & \left[\frac{1}{\bar{a}\left(\tau\right)}\,\frac{d\bar{a}\left(\tau\right)}{d\tau}\right]^{2}=\frac{\Omega_{m0}}{\bar{\phi}\left(\tau\right)}\left(\frac{1}{\bar{a}^{3}\left(\tau\right)}+\frac{V\left(\bar{\phi}\right)}{6\,m^{2}}\right)\nonumber \\
 & \quad-\frac{1}{\bar{a}\left(\tau\right)\,\bar{\phi}\left(\tau\right)}\,\frac{d\bar{a}\left(\tau\right)}{d\tau}\,\frac{d\bar{\phi}\left(\tau\right)}{d\tau}\,,\label{subeq:generalized-Friedmann-FLRW-tau}\\
 & \frac{1}{\bar{a}\left(\tau\right)}\,\frac{d^{2}\bar{a}\left(\tau\right)}{d\tau^{2}}=-\frac{\Omega_{m0}}{\bar{\phi}\left(\tau\right)}\left(\frac{1}{\bar{a}^{3}\left(\tau\right)}+\frac{V\left(\bar{\phi}\right)}{6\,m^{2}}\right)\nonumber \\
 & \quad+\frac{\Omega_{m0}}{6m^{2}}\frac{dV\left(\bar{\phi}\right)}{d\bar{\phi}}+\frac{1}{\bar{a}\left(\tau\right)\,\bar{\phi}\left(\tau\right)}\,\frac{d\bar{a}\left(\tau\right)}{d\tau}\,\frac{d\bar{\phi}\left(\tau\right)}{d\tau}\,,\label{subeq:generalized-acc-FLRW-tau}\\
 & \frac{3}{\bar{\phi}\left(\tau\right)}\,\frac{d^{2}\bar{\phi}\left(\tau\right)}{d\tau^{2}}-2\,\frac{\Omega_{m0}}{\bar{\phi}\left(\tau\right)}\,\frac{V\left(\bar{\phi}\right)}{m^{2}}+\frac{\Omega_{m0}}{m^{2}}\,\frac{dV\left(\bar{\phi}\right)}{d\bar{\phi}}\nonumber \\
 & \quad+\frac{9}{\bar{a}\left(\tau\right)\,\bar{\phi}\left(\tau\right)}\,\frac{d\bar{a}\left(\tau\right)}{d\tau}\,\frac{d\bar{\phi}\left(\tau\right)}{d\tau}=\frac{3\,\Omega_{m0}}{\bar{\phi}\left(\tau\right)\,\bar{a}^{3}\left(\tau\right)}\,.\label{subeq:scalar-field-FLRW-tau}
\end{align}
\end{subequations}
We used the usual relation for $\Omega_{m0}$, and we recall that $m^{2}=\chi\,\bar{\rho}/3$. In particular, Eqs.~\eqref{subeq:generalized-Friedmann-FLRW-tau} and \eqref{subeq:scalar-field-FLRW-tau} allow us to obtain numerical solutions for $\bar{a}\left(\tau\right)$ and $\bar{\phi}\left(\tau\right)$, while Eq.~\eqref{subeq:generalized-acc-FLRW-tau} is useful to estimate numerically the deceleration parameter $\bar{q}\left(\tau\right)$.

\begin{figure}
\centering\includegraphics[scale=0.24]{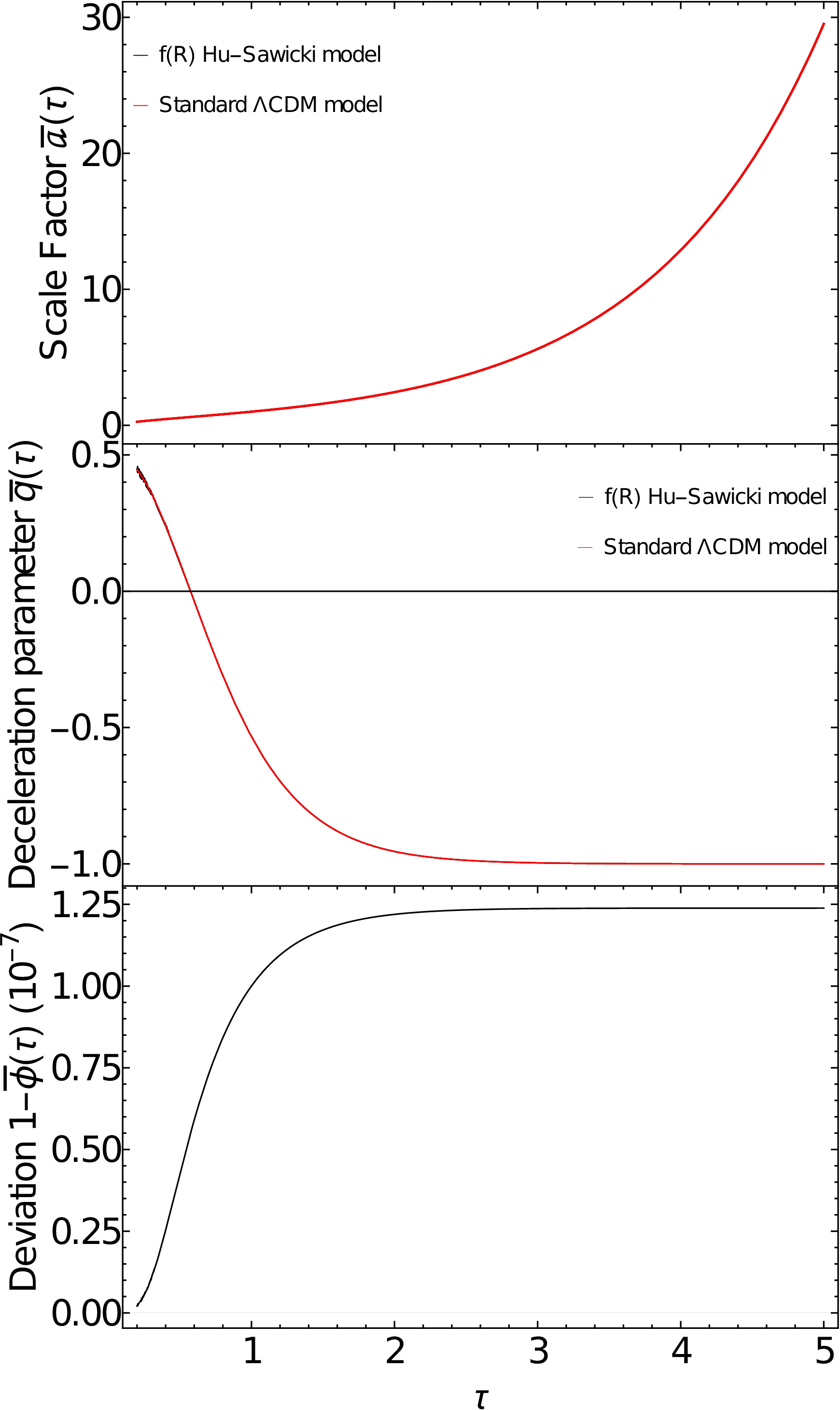}

\caption{Numerical background solutions, considering a flat FLRW geometry, for the $f\left(R\right)$ HS model in the Jordan frame. Top panel: evolution of the scale factor $\bar{a}$ in terms of the parameter $\tau$. Middle panel: behavior of the deceleration parameter $\bar{q}\left(\tau\right)$. Bottom panel: the deviation from the GR scenario ($\bar{\phi}=1$) with units of $10^{-7}$ for the vertical axis.}
\label{fig:HS-background}
\end{figure}

We choose the parameters of the model in such a way that the background modified gravity scenario is almost equivalent to the $\Lambda$CDM cosmological model with the aim of focusing later on the differences between the linear perturbation solutions in the two models. More precisely, we fix $\Omega_{m0}=0.3111,$ the same value adopted in \prettyref{subsec:Background-solution}, and we set the value $\left|F_{R0}\right|=1.0\times10^{-7}$ at the present cosmic time (redshift $z=0$ or $\tau=1$), which provides information about the deviation from the GR scenario, according to Eq.~\eqref{eq:JF-deviation-GR}. As a consequence, we constrain the HS dimensionless parameters: $c_{1}=2.0\times10^{6}$ and $c_{2}=1.5\times10^{5}$, as developed in \prettyref{subsec:The-Hu-Sawicki-model}. We recall that the profile of the background quantity $V\left(\bar{\phi}\right)/m^{2}$ is plotted in Fig.~\ref{fig:HS-background-potential}. 

To guarantee a nearly frozen evolution of the scalar field $\bar{\phi}$ for increasing $\tau$, we impose the following condition: $\frac{d\bar{\phi}}{d\tau}\left(\tau=1\right)=0$. Finally, we solve numerically Eqs.~\eqref{subeq:generalized-Friedmann-FLRW-tau} and \eqref{subeq:scalar-field-FLRW-tau} for $0.2<\tau<5$, when the relativistic components remain negligible as compared with the matter, to obtain the evolution of $\bar{a}\left(\tau\right)$ and $\bar{\phi}\left(\tau\right)$. The numerical results are shown in Fig.~\ref{fig:HS-background}. By comparing it with Fig.~\ref{fig:background-scale-factor-and-q}, it should be emphasized that $\bar{a}\left(\tau\right)$ and $\bar{q}\left(\tau\right)$ exhibit almost the same behavior in GR and in the Jordan frame of $f\left(R\right)$ gravity, as desired according to the choice of model parameters abovementioned. In particular, in the bottom panel of Fig.~\ref{fig:HS-background}, we plot the quantity $\left|1-\bar{\phi}\left(\tau\right)\right|$ to evaluate the deviation from GR ($\bar{\phi}=1$): we observe more relevant deviations in the late Universe for $\tau>1$, but nevertheless the background modified gravity dynamics still remains  almost undistinguishable from the $\Lambda$CDM scenario. Hence, we can shift the attention towards the first-order perturbation solutions in the Jordan frame of $f\left(R\right)$ gravity.

\subsection{Linearly perturbed solutions in an inhomogeneous Universe}\label{subsec:Linear-solution-JF}

We focus on the first-order perturbed equations to study the evolution of spherically symmetric perturbations. Considering the split between background terms and linear perturbations according to Eq.~\eqref{eq:expansion-background+pert-Jordan}, the set of field equations \eqref{subeqs:LTB-Jordan-frame} becomes
\begin{subequations}
\label{eq-linearized-JF-LTB}
\begin{align}
 & \delta\dot{\beta}^{\prime}=\frac{1}{r}\,\left(\delta\dot{\alpha}-\delta\dot{\beta}\right)-\frac{1}{2\,\bar{\phi}}\,\left(\delta\dot{\phi}^{\prime}-\frac{\dot{\bar{a}}}{\bar{a}}\,\delta\phi^{\prime}\right)\,,\label{subeq:linearized-01-LTB-JF}\\
 & \frac{2}{\bar{a}^{2}}\,\left[\frac{1}{r^{2}}\,\left(\delta\alpha-\delta\beta\right)-\delta\beta^{\prime\prime}-\frac{3}{r}\,\delta\beta^{\prime}+\frac{1}{r}\,\delta\alpha^{\prime}\right]\nonumber \\
 & \quad+\left(2\,\frac{\dot{\bar{a}}}{\bar{a}}+\frac{\dot{\bar{\phi}}}{\bar{\phi}}\right)\,\left(\delta\dot{\alpha}+2\,\delta\dot{\beta}\right)=\frac{\chi}{\bar{\phi}}\,\delta\rho-3\,\frac{\dot{\bar{a}}}{\bar{a}}\,\frac{\delta\dot{\phi}}{\bar{\phi}}\nonumber \\
 & \quad+\left(\frac{1}{2}\,\left.\frac{dV}{d\phi}\right|_{\phi=\bar{\phi}}-\frac{1}{2}\,\frac{V\left(\bar{\phi}\right)}{\bar{\phi}}-\frac{\chi\,\bar{\rho}}{\bar{\phi}}+3\,\frac{\dot{\bar{a}}}{\bar{a}}\,\frac{\dot{\bar{\phi}}}{\bar{\phi}}\right)\,\frac{\delta\phi}{\bar{\phi}}\nonumber \\
 & \quad+\frac{1}{\bar{a}^{2}\,\bar{\phi}}\,\left(\delta\phi^{\prime\prime}+\frac{2}{r}\,\delta\phi^{\prime}\right)\,,\label{subeq:linearized-00-LTB-JF}\\
 & \delta\ddot{\beta}+3\,\frac{\dot{\bar{a}}}{\bar{a}}\,\delta\dot{\beta}+\frac{1}{\bar{a}^{2}\,r^{2}}\,\left(\delta\alpha-\delta\beta-r\,\delta\beta^{\prime}\right)=\nonumber \\
 & \quad\left(\frac{1}{2}\,\left.\frac{dV}{d\phi}\right|_{\phi=\bar{\phi}}-\frac{1}{2}\,\frac{V\left(\bar{\phi}\right)}{\bar{\phi}}+\frac{\ddot{\bar{\phi}}}{\bar{\phi}}+2\,\frac{\dot{\bar{a}}}{\bar{a}}\,\frac{\dot{\bar{\phi}}}{\bar{\phi}}\right)\,\frac{\delta\phi}{2\,\bar{\phi}}\nonumber \\
 & \quad-\frac{1}{2\,\bar{\phi}}\,\left(\delta\ddot{\phi}+2\,\frac{\dot{\bar{a}}}{\bar{a}}\,\delta\dot{\phi}+2\,\dot{\bar{\phi}}\,\delta\dot{\beta}-\frac{2}{\bar{a}^{2}\,r}\,\delta\phi^{\prime}\right)\,.\label{subeq:linearized-11-LTB-JF}
\end{align}
\end{subequations}

\noindent Similarly, starting from Eq.~\eqref{eq: scalar LTB Jordan frame}, the linearized scalar field equation is given by
\begin{align}
 & \delta\ddot{\phi}+3\frac{\dot{\bar{a}}}{\bar{a}}\delta\dot{\phi}+\left(\delta\dot{\alpha}+2\delta\dot{\beta}\right)\dot{\bar{\phi}}-\frac{1}{\bar{a}^{2}}\left(\delta\phi^{\prime\prime}+\frac{2}{r}\delta\phi^{\prime}\right)\nonumber \\
 & \quad+\frac{1}{3}\,\left(\bar{\phi}\,\left.\frac{d^{2}V}{d\phi^{2}}\right|_{\phi=\bar{\phi}}-\left.\frac{dV}{d\phi}\right|_{\phi=\bar{\phi}}\right)\,\delta\phi=\frac{1}{3}\,\chi\,\delta\rho\,.\label{eq:linearized-scalar-field-LTB-JF}
\end{align}

\noindent Moreover, the continuity equation~\eqref{eq:continuity-LTB-general-metric} rewrites as
\begin{equation}
\delta\dot{\rho}+3\,\frac{\dot{\bar{a}}}{\bar{a}}\,\delta\rho+\left(\delta\dot{\alpha}+2\,\delta\dot{\beta}\right)\,\bar{\rho}=0\label{eq:linearized-contin-LTB-JF}
\end{equation}
at the linear perturbation order. We have used Eq.~\eqref{eq:alphabar(t)-betabar(t)-a(t)} to rewrite $\bar{\alpha}$ and $\bar{\beta}$ in terms of $\bar{a}$. 

Following the same approach we adopted in GR in \prettyref{sec:pert-approach-GR}, we use a separation of variables method to study separately the evolution of inhomogeneities in time and space. Hence, we factorize the linear perturbations as:
\begin{align}
\delta\alpha\left(t,r\right) & \equiv A_{p}\left(t\right)\,\mathcal{A}_{p}\left(r\right)\nonumber\\
\delta\beta\left(t,r\right) & \equiv B_{p}\left(t\right)\,\mathcal{B}_{p}\left(r\right)\nonumber \\
\delta\rho\left(t,r\right) & \equiv P_{p}\left(t\right)\,\varrho_{p}\left(r\right)\nonumber \\
\delta\phi\left(t,r\right) & \equiv\Phi_{p}\left(t\right)\,\varphi_{p}\left(r\right)\,.\label{eq:separation-variables-LTB-JF}
\end{align}

Assuming this factorization, we can rewrite Eqs.~\eqref{eq-linearized-JF-LTB}, \eqref{eq:linearized-scalar-field-LTB-JF}, and \eqref{eq:linearized-contin-LTB-JF}. However, we notice the presence of several mixed terms depending both on $t$ and $r$, which do not allow us to solve the equations using the separation of variables in a standard way, unless we rely on reasonable and simplifying assumptions (further details on explicit calculations are in the Appendix~\ref{app:technical}). For instance, we are able to use the separation of variables method for all field equations, if we require the two following conditions:
\begin{subequations}
\label{eq-ipotesi-semplificative}
\begin{align}
A_{p} & =\lambda_{1}\,B_{p}\,,\label{subeq:ipotesi-semplificativa1}\\
\varrho_{p} & =\lambda_{2}\,\varphi_{p}\,,\label{subeq:ipotesi-semplificativa2}
\end{align}
\end{subequations}
where $\lambda_{1}$ and $\lambda_{2}$ are two proportionality constants. These conditions allow us to simplify the equation system and easily separate time and radial dependences: we obtain a set of differential equations describing the radial profiles of perturbations and another equation system concerning only the time evolution. 

Hence, starting from Eqs.~\eqref{eq-linearized-JF-LTB}, \eqref{eq:linearized-scalar-field-LTB-JF}, and \eqref{eq:linearized-contin-LTB-JF}, after long but straightforward calculations (see the Appendix \ref{app:technical}), we obtain a set of equations for the radial part:
\begin{subequations}
\label{eq-linearized-JF-LTB-separated-only-r}
\begin{align}
 & \mathcal{A}_{p}=\frac{1}{\lambda_{1}}\,\left[\mathcal{B}_{p}+r\,\left(\mathcal{B}_{p}^{\prime}+\mu_{1}\,\varphi_{p}^{\prime}\right)\right]\,,\label{subeq:linearized-01-LTB-JF-split-r}\\
 & \mathcal{B}_{p}=\frac{2}{\mu_{4}\,r}\,\varphi_{p}^{\prime}-\mu_{1}\,\varphi_{p}\,,\label{subeq:linearized-11-LTB-JF-split-r}\\
 & \varphi_{p}^{\prime\prime}+\frac{2}{r}\,\varphi_{p}^{\prime}-\mu_{3}^{2}\,\varphi_{p}=0\,,\label{subeq:linearized-scalar-field-LTB-JF-split-r}\\
 & \varrho_{p}=\frac{1}{\mu_{2}}\,\left(\lambda_{1}\,\mathcal{A}_{p}+2\,\mathcal{B}_{p}\right)\,,\label{subeq:linearized-contin-LTB-JF-split-r}
\end{align}
\end{subequations}
where $\mu_{1}$, $\mu_{2}$, $\mu_{3}$, and $\mu_{4}$ are constants arising from the separation of variables. We have four unknown quantities ($\mathcal{A}_{p}$, $\mathcal{B}_{p}$, $\varrho_{p}$, and $\varphi_{p}$), which are related through the condition given by Eq.~\eqref{subeq:ipotesi-semplificativa2}, the perturbed 0-1 component of field equations~\eqref{subeq:linearized-01-LTB-JF-split-r}, the perturbed 1-1 component~\eqref{subeq:linearized-11-LTB-JF-split-r}, the linearized scalar field equation~\eqref{subeq:linearized-scalar-field-LTB-JF-split-r}, and in addition the perturbed continuity equation~\eqref{subeq:linearized-contin-LTB-JF-split-r}. In particular, note that $\mu_{3}$ has dimensions of reciprocal length, i.e., $\left[\mu_{3}\right]=L^{-1}$, as you can see from Eq.~\eqref{subeq:linearized-scalar-field-LTB-JF-split-r}.

Similarly, in the Appendix~\ref{app:technical}, we write an equation system for the time evolution of perturbations:
\begin{subequations}
\label{eq-linearized-JF-LTB-separated-only-t}
\begin{align}
 & \dot{B}_{p}=\frac{1}{2\,\bar{\phi}\,\mu_{1}}\,\left(\dot{\Phi}_{p}-\frac{\dot{\bar{a}}}{\bar{a}}\,\Phi_{p}\right)\,,\label{subeq:linearized-01-LTB-JF-split-t}\\
 & \frac{\mu_{2}\lambda_{2}}{2\mu_{1}}\left(2\frac{\dot{\bar{a}}}{\bar{a}}+\frac{\dot{\bar{\phi}}}{\bar{\phi}}\right)\left(\frac{\dot{\Phi}_{p}}{\Phi_{p}}-\frac{\dot{\bar{a}}}{\bar{a}}\right)+\frac{\chi\bar{\rho}}{\bar{\phi}}-\frac{1}{2}\left.\frac{dV}{d\phi}\right|_{\phi=\bar{\phi}}\nonumber \\
 & \quad+\frac{V\left(\bar{\phi}\right)}{2\bar{\phi}}+3\frac{\dot{\bar{a}}}{\bar{a}}\left(\frac{\dot{\Phi}_{p}}{\Phi_{p}}-\frac{\dot{\bar{\phi}}}{\bar{\phi}}\right)-\chi\lambda_{2}\frac{P_{p}}{\Phi_{p}}\nonumber\\
 & \quad+2\mu_{1}\mu_{3}^{2}\frac{\bar{\phi}B_{p}}{\bar{a}^{2}\Phi_{p}}=\frac{\mu_{3}^{2}}{\bar{a}^{2}}\,,\label{subeq:linearized-00-LTB-JF-split-t}\\
 & \ddot{B}_{p}+\dot{B}_{p}\,\left(3\,\frac{\dot{\bar{a}}}{\bar{a}}+\frac{\dot{\bar{\phi}}}{\bar{\phi}}\right)=\frac{\mu_{4}}{2\,\bar{a}^{2}}\,\left(\frac{\Phi_{p}}{\bar{\phi}}-\mu_{1}\,B_{p}\right)\,,\label{subeq:linearized-11-LTB-JF-split-t}\\
 & \frac{\ddot{\Phi}_{p}}{\Phi_{p}}+3\,\frac{\dot{\bar{a}}}{\bar{a}}\,\frac{\dot{\Phi}_{p}}{\Phi_{p}}-\frac{1}{3}\,\left.\frac{dV}{d\phi}\right|_{\phi=\bar{\phi}}+\frac{1}{3}\,\bar{\phi}\,\left.\frac{d^{2}V}{d\phi^{2}}\right|_{\phi=\bar{\phi}}\nonumber \\
 & \quad+\mu_{2}\,\lambda_{2}\,\dot{\bar{\phi}}\,\frac{\dot{B}_{p}}{\Phi_{p}}-\frac{\chi\,\lambda_{2}}{3}\,\frac{P_{p}}{\Phi_{p}}=\frac{\mu_{3}^{2}}{\bar{a}^{2}}\,,\label{subeq:linearized-scalar-field-LTB-JF-split-t}\\
 & \dot{P}_{p}+3\,\frac{\dot{\bar{a}}}{\bar{a}}\,P_{p}+\mu_{2}\,\bar{\rho}\,\dot{B}_{p}=0\,.\label{subeq:linearized-contin-LTB-JF-split-t}
\end{align}
\end{subequations}

We have four unknown quantities ($A_{p}$, $B_{p}$, $P_{p}$, and $\Phi_{p}$) for the time evolution of inhomogeneities, which are fully described by the assumption~\eqref{subeq:ipotesi-semplificativa1}, the perturbed 0-1, 0-0, 1-1 components of field equations given by Eqs.~\eqref{subeq:linearized-01-LTB-JF-split-t}, \eqref{subeq:linearized-00-LTB-JF-split-t}, \eqref{subeq:linearized-11-LTB-JF-split-t}, respectively, the linearized scalar field equation~\eqref{subeq:linearized-scalar-field-LTB-JF-split-t}, and the perturbed continuity equation~\eqref{subeq:linearized-contin-LTB-JF-split-t}. Note that two equations of the latter list are redundant since the scalar field and continuity equations are not independent of the other field equations. 

It should be stressed that the potential $V\left(\phi\right)$ affects only the time evolution. As a consequence, since the scalar field potential is related to a specific modified $f\left(R\right)$ model, the time evolution strongly depends on the particular modified gravity model considered, while the radial part of the linearized equations is completely model free. This is a crucial point to identify a peculiar feature of the inhomogeneities evolution in the Jordan frame of $f\left(R\right)$ gravity through analysis of the radial profiles of perturbations. 

Once we have split all field equations in time and radial contributions, we seek linear order solutions separately.

\subsubsection{Radial profiles}\label{subsec:Radial-solutions-JF}

If we focus on the radial evolution of perturbations, we can solve the respective equation system analytically. More specifically, in solving the differential equation~\eqref{subeq:linearized-scalar-field-LTB-JF-split-r}, we obtain the Yukawa behavior for the radial solution of the linearly perturbed scalar field:
\begin{equation}
\varphi_{p}\left(r\right)=\frac{\gamma}{r}\,e^{-\mu_{3}\,r}\,.\label{eq:Yukawa-solution-radial-scalar-field-pert}
\end{equation}
We have set perturbations to vanish at infinity according to the cosmological principle, hence we have only one integration constant $\gamma$. We recall that $\left[\mu_{3}\right]=L^{-1}$, as it can be checked also from Eq.~\eqref{eq:Yukawa-solution-radial-scalar-field-pert}. Note that $\varrho_{p}\left(r\right)$ has the same radial dependence, i.e.
\begin{equation}
\varrho_{p}\left(r\right)=\frac{\lambda_{2}\,\gamma}{r}\,e^{-\mu_{3}\,r}\,,\label{eq:Yukawa-solution-radial-matter-pert}
\end{equation}
in which we considered the assumption given by Eq.~\eqref{subeq:ipotesi-semplificativa2}. 

Consequently, from Eq.~\eqref{subeq:linearized-11-LTB-JF-split-r}, we also get
\begin{equation}
\mathcal{B}_{p}\left(r\right)=-\frac{\gamma}{\mu_{4}}\,e^{-\mu_{3}\,r}\,\left(\frac{\mu_{4}\,\mu_{1}}{r}+\frac{2\,\mu_{3}^{2}}{r^{2}}+\frac{2}{r^{3}}\right)\,,\label{eq:Yukawa-solution-radial-B-pert}
\end{equation}
where we used the solution~\eqref{eq:Yukawa-solution-radial-scalar-field-pert}. Finally, we combine Eqs.~\eqref{subeq:linearized-01-LTB-JF-split-r}, \eqref{eq:Yukawa-solution-radial-scalar-field-pert}, and \eqref{eq:Yukawa-solution-radial-B-pert} to obtain the scalar perturbation
\begin{equation}
\mathcal{A}_{p}\left(r\right)=\frac{\gamma}{\lambda_{1}}e^{-\mu_{3}r}\left(\frac{2\mu_{1}+\mu_{2}\lambda_{2}}{r}+\frac{4\mu_{3}^{2}}{\mu_{4}r^{2}}+\frac{4}{\mu_{4}r^{3}}\right)\,.\label{eq:Yukawa-solution-radial-A-pert}
\end{equation}

It should be noted that, in order to satisfy the compatibility between Eqs.~\eqref{subeq:linearized-01-LTB-JF-split-r}, \eqref{subeq:ipotesi-semplificativa2}, \eqref{subeq:linearized-scalar-field-LTB-JF-split-r}, \eqref{subeq:linearized-11-LTB-JF-split-r}, and \eqref{subeq:linearized-contin-LTB-JF-split-r}, we require the following relation
\begin{equation}
\mu_{4}=\frac{2\,\mu_{3}^{2}}{3\,\mu_{1}+\mu_{2}\,\lambda_{2}}\label{eq:condizione-costanti-compatibilit=0000E0}
\end{equation}
between the constants involved in the set of equations\footnote{More specifically, to write this compatibility condition, we have started from Eqs.~\eqref{subeq:linearized-01-LTB-JF-split-r}, \eqref{subeq:linearized-contin-LTB-JF-split-r}, and then we have replaced $\varrho_{p}$ and $\mathcal{B}_{p}$ with $\varphi_{p}$ and $\varphi_{p}^{\prime}$ by using Eqs.~\eqref{subeq:ipotesi-semplificativa2} and \eqref{subeq:linearized-11-LTB-JF-split-r}, respectively; finally, the resulting equation is compared with Eq.~\eqref{subeq:linearized-scalar-field-LTB-JF-split-r} to set the above relation~\eqref{eq:condizione-costanti-compatibilit=0000E0} between the constants.}. In particular, note that $\left[\mu_{4}\right]=L^{-2}$, since $\mu_{3}$ has the dimension of inverse length, and the other constants are dimensionless. 

It should be pointed out that the $f\left(R\right)$ gravity establishes a typical radial scale, $r_{c}\equiv\mu_{3}^{-1}$, such that local inhomogeneities vanish for $r\gg r_{c}$ more rapidly than the respective perturbations in the $\Lambda$LTB model. For instance, this fact can be shown by comparing the behavior of the perturbed scale factor given in Eq.~\eqref{eq:soluz-pert-scale-factor-GR} with the radial profiles of the LTB metric functions in Eqs.~\eqref{eq:Yukawa-solution-radial-B-pert} and \eqref{eq:Yukawa-solution-radial-A-pert}. It is important to stress that our result is completely independent of the choice of scalar field potential $V\left(\bar{\phi}\right)$. Hence these radial solutions apply to any $f(R)$ extended model, providing a remarkable feature of the radial evolution within the Jordan frame of the $f\left(R\right)$ gravity as compared to GR. 

\subsubsection{Time evolution}\label{subsec:Time-solutions-JF}

Now we focus on the time evolution of perturbations in the Jordan frame. Clearly, the choice of the background $f\left(R\right)$ modified gravity model affects the time dependence of inhomogeneities since $V\left(\bar{\phi}\right)$ appears in almost all equations regarding the time part. Our main result concerns the peculiarity of the radial profiles of perturbations in the Jordan frame, which does not depend on a specific modified gravity model. Here, we merely want to prove the existence of at least one stable time solution. 

To solve numerically the equation system given by Eqs.~\eqref{subeq:ipotesi-semplificativa1}, \eqref{subeq:linearized-01-LTB-JF-split-t}, \eqref{subeq:linearized-contin-LTB-JF-split-t}, \eqref{subeq:linearized-scalar-field-LTB-JF-split-t}, \eqref{subeq:linearized-00-LTB-JF-split-t}, \eqref{subeq:linearized-11-LTB-JF-split-t}, we rewrite it in terms of the dimensionless parameter $\tau$, defined in Eq.~\eqref{eq:definition_tau}. We also recall that $d/dt\approx H_{0}\,d/d\tau$. 

In particular, Eqs.~\eqref{subeq:ipotesi-semplificativa1} and \eqref{subeq:linearized-01-LTB-JF-split-t} maintain the same form if we write them in terms of $\tau$ as 
\begin{align}
A_{p}\left(\tau\right) & =\lambda_{1}\,B_{p}\left(\tau\right)\,,\label{eq:ipotesi-semplificativa-tau}\\
\frac{dB_{p}}{d\tau} & =\frac{1}{2\,\bar{\phi}\left(\tau\right)\,\mu_{1}}\,\left(\frac{d\Phi_{p}}{d\tau}-\frac{1}{\bar{a}}\,\frac{d\bar{a}}{d\tau}\,\Phi_{p}\left(\tau\right)\right)\,,\label{eq:linearized-01-LTB-JF-split-t-tau}
\end{align}
respectively. 

\noindent Furthermore, the linearized scalar field equation \eqref{subeq:linearized-scalar-field-LTB-JF-split-t} becomes
\begin{align}
 & \frac{1}{\Phi_{p}\left(\tau\right)}\,\frac{d^{2}\Phi_{p}}{d\tau^{2}}+3\,\frac{1}{\bar{a}\left(\tau\right)}\,\frac{d\bar{a}}{d\tau}\,\frac{1}{\Phi_{p}\left(\tau\right)}\,\frac{d\Phi_{p}}{d\tau}\nonumber \\
 & \quad-\frac{1}{3}\frac{\Omega_{m0}}{m^{2}}\left.\frac{dV}{d\phi}\right|_{\phi=\bar{\phi}\left(\tau\right)}+\frac{1}{3}\bar{\phi}\left(\tau\right)\frac{\Omega_{m0}}{m^{2}}\left.\frac{d^{2}V}{d\phi^{2}}\right|_{\phi=\bar{\phi}\left(\tau\right)}\nonumber \\
 & \quad+\mu_{2}\lambda_{2}\frac{d\bar{\phi}}{d\tau}\frac{1}{\Phi_{p}\left(\tau\right)}\frac{dB_{p}}{d\tau}-\frac{\lambda_{2}\Omega_{m0}\Xi\left(\tau\right)}{\bar{a}^{3}\left(\tau\right)\Phi_{p}\left(\tau\right)}=\frac{\tilde{\mu}_{3}^{2}}{\bar{a}^{2}\left(\tau\right)}\,.\label{eq:linearized-scalar-field-LTB-JF-split-t-tau}
\end{align}
We have defined the density contrast $\Xi\left(\tau\right)\equiv P_{p}\left(\tau\right)/\bar{\rho}\left(\tau\right)$ for linear perturbations. Note also that the quantity $\tilde{\mu}_{3}\equiv\mu_{3}/H_{0}$ is dimensionless in natural units. 

Similarly, Eq.~\eqref{subeq:linearized-00-LTB-JF-split-t} in terms of $\tau$ rewrites as
\begin{align}
 & \frac{\mu_{2}\lambda_{2}}{2\mu_{1}}\left(\frac{2}{\bar{a}\left(\tau\right)}\frac{d\bar{a}}{d\tau}+\frac{1}{\bar{\phi}\left(\tau\right)}\frac{d\bar{\phi}}{d\tau}\right)\left(\frac{1}{\Phi_{p}\left(\tau\right)}\frac{d\Phi_{p}}{d\tau}\right.\nonumber\\
 & \left.-\frac{1}{\bar{a}\left(\tau\right)}\frac{d\bar{a}}{d\tau}\right) + \frac{3\,\Omega_{m0}}{\bar{\phi}\left(\tau\right)\,\bar{a}^{3}\left(\tau\right)}\nonumber\\
 & +\frac{\Omega_{m0}}{2\,m^{2}}\left(\frac{V\left(\bar{\phi}\right)}{\bar{\phi}\left(\tau\right)}-\left.\frac{dV}{d\phi}\right|_{\phi=\bar{\phi}\left(\tau\right)}\right)+\frac{3}{\bar{a}\left(\tau\right)}\frac{d\bar{a}}{d\tau}\nonumber \\
 & \times\left(\frac{1}{\Phi_{p}\left(\tau\right)}\frac{d\Phi_{p}}{d\tau}-\frac{1}{\bar{\phi}\left(\tau\right)}\frac{d\bar{\phi}}{d\tau}\right)-\frac{3\lambda_{2}\Omega_{m0}\Xi\left(\tau\right)}{\Phi_{p}\left(\tau\right)\,\bar{a}^{3}\left(\tau\right)}\nonumber \\
 & +2\mu_{1}\tilde{\mu}_{3}^{2}\,\frac{B_{p}\left(\tau\right)}{\Phi_{p}\left(\tau\right)}\,\frac{\bar{\phi}\left(\tau\right)}{\bar{a}^{2}\left(\tau\right)}=\frac{\tilde{\mu}_{3}^{2}}{\bar{a}^{2}\left(\tau\right)}\,,\label{eq:00-LTB-JF-split-t-tau}
\end{align}
and Eq.~\eqref{subeq:linearized-11-LTB-JF-split-t} turns into:
\begin{align}
 & \frac{d^{2}B_{p}}{d\tau^{2}}+\frac{dB_{p}}{d\tau}\,\left(\frac{3}{\bar{a}\left(\tau\right)}\,\frac{d\bar{a}}{d\tau}+\frac{1}{\bar{\phi}\left(\tau\right)}\,\frac{d\bar{\phi}}{d\tau}\right)=\nonumber \\
 & \qquad\quad=\frac{\tilde{\mu}_{4}}{2\,\bar{a}^{2}\left(\tau\right)}\,\left(\frac{\Phi_{p}\left(\tau\right)}{\bar{\phi}\left(\tau\right)}-\mu_{1}\,B_{p}\left(\tau\right)\right)\,.\label{eq:linearized-11-LTB-JF-split-t-tau}
\end{align}
In the latter equation, we have introduced the dimensionless quantity $\tilde{\mu}_{4}\equiv\mu_{4}/H_{0}^{2}$ in natural units.

\begin{figure}[ht]
\centering \includegraphics[scale=0.26]{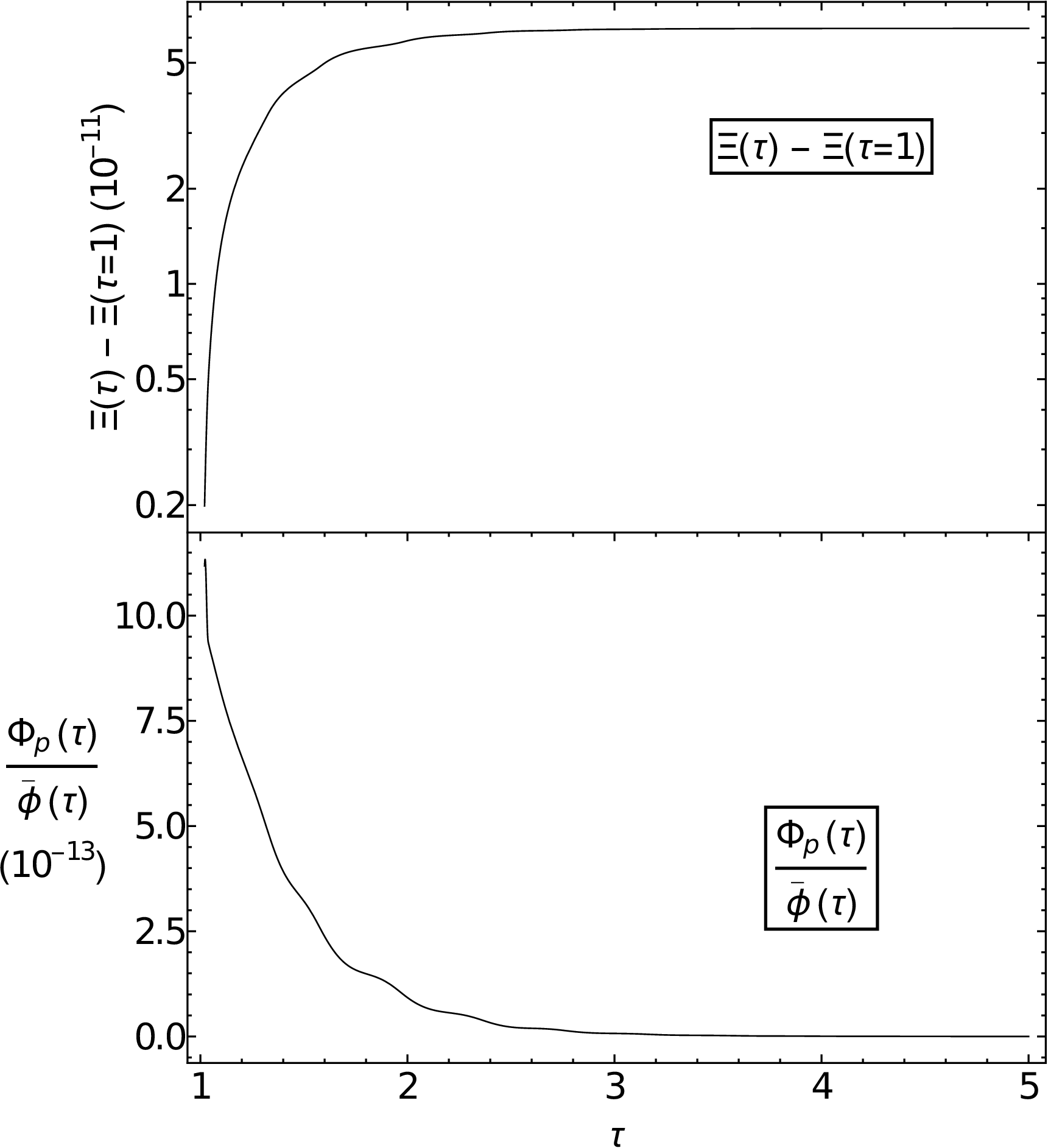}

\caption{Numerical solutions of linear perturbations in the Jordan frame of $f\left(R\right)$ gravity in terms of the parameter $\tau$. Top panel: the difference between the density contrast $\Xi$ evaluated at a generic $\tau$ and $\tau=1$ today represented in units of $10^{-11}$ and logarithmic scale. Bottom panel: the ratio between the linearly perturbed scalar field $\Phi_{p}\left(\tau\right)$ and the respective background scalar field $\bar{\phi}\left(\tau\right)$ in units of $10^{-13}$.}
\label{fig:HS-pert-density-contrast}
\end{figure}

Finally, considering the definition of the density contrast $\Xi\left(\tau\right)$ and the background equation~\eqref{eq:continuity-background-LCDM}, it is straightforward to show that the linearly perturbed continuity equation~\eqref{subeq:linearized-contin-LTB-JF-split-t} simply rewrites in terms of $\tau$ as
\begin{equation}
\frac{d\Xi}{d\tau}+\mu_{2}\,\frac{dB_{p}}{d\tau}=0\,.\label{eq:linearized-contin-LTB-JF-split-t-tau}
\end{equation}
Note that we can use Eq.~\eqref{eq:linearized-contin-LTB-JF-split-t-tau} to reduce the number of unknown quantities in the other field equations~\eqref{eq:linearized-scalar-field-LTB-JF-split-t-tau}. 

Now we want to solve numerically Eqs.~\eqref{eq:linearized-01-LTB-JF-split-t-tau} and \eqref{eq:linearized-scalar-field-LTB-JF-split-t-tau} to obtain $\Xi\left(\tau\right)$ and $\Phi_{p}\left(\tau\right)$. In this regard, we fix the values of constants for simplicity $\mu_{1}=1$, $\mu_{2}=10^{2}$, $\lambda_{2}=5$, $\tilde{\mu}_{3}=1$ and set initial conditions at $\tau=1$ for inhomogeneities: $\Xi\left(1\right)=10^{-5}$, $\Phi_{p}\left(1\right)=10^{-12}$, and $d\Phi_{p}/d\tau\left(1\right)=0$. These values have been chosen to have weak and slowly varying perturbations and guarantee the existence of a stable numerical solution. Moreover, we use the background numerical solutions for $\bar{a}\left(\tau\right)$ and $\bar{\phi}\left(\tau\right)$, and the scalar field potential $V\left(\bar{\phi}\right)$ for the HS model discussed in \prettyref{subsec:Background-solution-JF}. 

Then, we obtain numerical solutions for $\Xi\left(\tau\right)$ and $\Phi_{p}\left(\tau\right)$, which are plotted in Fig.~\ref{fig:HS-pert-density-contrast}. It should be noted that these solutions are stable in time, since inhomogeneous perturbations are dominated by background terms for any $\tau$. 

Once we have obtained $\Xi$ and $\Phi_{p}$ numerically, the remaining perturbed quantities $A_{p}$ and $B_{p}$ can be easily found by using Eqs.~\eqref{eq:ipotesi-semplificativa-tau} and \eqref{eq:linearized-contin-LTB-JF-split-t-tau}.

\section{Conclusions}\label{sec:conclusions}

In this paper, we have analyzed the LTB spherically symmetric solution, linearized over a flat FLRW background, comparing its morphology in GR and $f\left(R\right)$ modified gravity theories, as viewed in the Jordan frame. In the former case, we have referred to the $\Lambda$LTB model, including a matter fluid and a cosmological constant; in the latter model, we have considered the $f\left(R\right)$ Hu-Sawicki formulation for the dark energy component of the background Universe, in which the cosmic acceleration is driven by the no-Einsteinian geometrical terms if compared it to GR. 

We have studied the dynamics in both cosmological scenarios to highlight the peculiarities of these models. To describe spherically symmetric deviations from homogeneity in the late Universe, we have used the separation of variables method to address the partial differential equations system for the first-order perturbation. Then, the radial component of such a reduction procedure has been analytically integrated, while the-time dependent part has required a numerical treatment. 

The key difference between the two cosmological models studied in this work concerns the different form of the 0-1 component of the gravitational field equations. Indeed, in the $\Lambda$LTB model, this equation~\eqref{subeq:LTB-01-GR} can be easily solved by removing one of the two free metric functions of the problem, hence the LTB metric takes the well-known simplified form~\eqref{eq:LTBmetric-GR-simpler}. On the other hand, within the framework of the $f\left(R\right)$ gravity in the Jordan frame, the 0-1 component given by Eq.~\eqref{subeq: 01 LTB Jordan frame} intrinsically links the metric tensor components to the non-minimally coupled scalar field. Thus, the non-minimally coupling prevents the simplification above unlike GR, and we have to deal with two distinct metric functions $\alpha\left(t,r\right)$ and $\beta\left(t,r\right)$ in the LTB metric given by Eq.~\eqref{eq:LTBmetric}.

Concerning the radial solutions, GR provides only a natural decay of the perturbations for large $r$ values, following a power law, as it emerges from Eq.~\eqref{eq:soluz-pert-scale-factor-GR}. Differently, in the $f\left(R\right)$ modified gravity formulation, we obtain the Yukawa-like decaying for the radial perturbations of the scalar functions given by Eqs.~\eqref{eq:Yukawa-solution-radial-scalar-field-pert}, \eqref{eq:Yukawa-solution-radial-matter-pert}, \eqref{eq:Yukawa-solution-radial-B-pert}, and \eqref{eq:Yukawa-solution-radial-A-pert}. Furthermore, we have employed the $f\left(R\right)$ HS model to describe the background Universe for comparison with the $\Lambda$CDM model, but we could have considered other viable $f\left(R\right)$ modified gravity models. Indeed, we stress that our main result regarding the Yukawa-like radial perturbations does not depend on the $f\left(R\right)$ functional form. 

It should be noted that Yukawa-like radial profiles are recurring in $f\left(R\right)$ theories, as evidenced, for instance, in other studies \citep{Sanders1984,Capozziello:2012ie,Stabile:2013jon,deAlmeida:2018kwq} by the presence of Yukawa-like corrections in the Newton potential with consequent implications for the dark matter problem and the flat rotation curves of galaxies. 

In this paper, the different morphology of the radial solutions between the $\Lambda$LTB model and inhomogeneous $f\left(R\right)$ cosmology must be regarded as the most relevant signature we fixed about the possibility to adopt the $f\left(R\right)$ modified gravity scenario to describe the accelerating late Universe via spherically symmetric perturbations over a homogeneous background. In other words, the different radial profiles of local inhomogeneities in the Universe may be possible hints of a theory beyond GR. 

For what concerned the time evolution of inhomogeneous perturbations both in the $\Lambda$LTB model and in the LTB solution as emerging from the $f\left(R\right)$ gravity in the Jordan frame, the numerical analysis has allowed us to outline that it is always possible to obtain a non-divergent amplitude of the perturbations as time goes by, according to the reliable idea of a stable homogeneous and isotropic Universe in the near future. 

Nevertheless, we are aware that our work has one limitation: the obtained radial solutions clearly diverge in the center of the LTB symmetry, where the observer, i.e., human location, is intended to be set. This feature simply suggests that our solution has a non-perturbative extension from a given large enough radial coordinate up to $r=0$, which is an important task for future investigations in the late Universe. The present analysis fosters further studies about inhomogeneous cosmology when regarded as a local (non-linear) deformation of the FLRW geometry, as it may be also possibly pointed out by local measurements of the Hubble constant.

The present study may have a relevant impact on the observations of the large-scale structure of the Universe, when forthcoming missions, like the Euclid satellite \citep{Amendola:2016saw}, detect the clumpy galaxy distribution with greater accuracy. The comparison between the $\Lambda$CDM model and $f\left(R\right)$ modified gravity theories through the investigation of spatial inhomogeneities may become a powerful tool to test the robustness of the cosmological concordance model. 

\backmatter





\section*{Declarations}

\textbf{Acknowledgments.}
We would like to thank Paolo Marcoccia for the contribution \citep{Marcoccia:2018anj} given to this subject in GR during his Master Degree thesis.

\textbf{Funding.} The work of TS is supported by the Galileo Galilei Institute Boost Fellowship and  was supported by the Della Riccia foundation grant for the year 2023. 

\textbf{Data availability statement.} This manuscript has no associated data. This is a theoretical work and no experimental data has been deposited.

\textbf{Code availability.} Code/Software sharing not applicable to this article as no code/software was generated or analysed during the current study.

\textbf{Conflict of interest.} The authors have no competing interests to declare that are relevant to the content of this article.


\begin{appendices}

\section{}\label{app:Verifying-spatial-isotropy}

We want to check that the spatial isotropy in the LTB geometry is also preserved by the field equations in the Jordan frame of $f\left(R\right)$ modified gravity. In particular, we prove that the 2-2 component of gravitational field equations depends on the other ones. Moreover, it can be easily checked that the 2-2 and 3-3 components are exactly the same. 

It should be noted that the isotropy in GR can be shown in field equations by employing Eq.~\eqref{eq:relazioneG11-G22-GR}. However, within the framework of the Jordan frame of $f\left(R\right)$ gravity, this simple relation no longer applies, because of the presence of the non-minimal coupling between the scalar field $\phi$ and the metric. Then, we need a careful analysis of field equations to verify the isotropy in the LTB metric. In this regard, we rewrite the gravitational field equations~\eqref{subeq: field equations Jordan frame - metric} in the Jordan frame as $G_{\mu\nu}=S_{\mu\nu}$, where
\begin{equation}
S_{\mu\nu}=\frac{\chi}{\phi}T_{\mu\nu}-\frac{1}{2\phi}g_{\mu\nu}V\left(\phi\right)+\frac{1}{\phi}\left(\nabla_{\mu}\nabla_{\nu}\phi-g_{\mu\nu}\boxempty\phi\right)
\end{equation}
denotes all the source elements in a compact way. 

Considering a pressure-less dust, the 2-2 component of Eqs.~\eqref{subeq: field equations Jordan frame - metric} in the Jordan frame in the LTB metric \eqref{eq:LTBmetric}, i.e., $G_{\,2}^{2}=S_{\,2}^{2}$, is written as
\begin{align}
 & \ddot{\alpha}+\ddot{\beta}+\dot{\alpha}^{2}+\dot{\beta}\left(\dot{\alpha}+\dot{\beta}\right)\nonumber\\
 & -e^{-2\alpha}\left[\beta^{\prime\prime}-\beta^{\prime}\left(\alpha^{\prime}-\beta^{\prime}\right)\right]=\frac{V}{2\phi}\nonumber\\
 & -\frac{1}{\phi}\left\{ \ddot{\phi}+\left(\dot{\alpha}+\dot{\beta}\right)\dot{\phi}-e^{-2\alpha}\left[\phi^{\prime\prime}-\phi^{\prime}\left(\alpha^{\prime}-\beta^{\prime}\right)\right]\right\} .\label{eq: 22 LTB Jordan frame-app}
\end{align}

\noindent For the sake of convenience, we also rewrite here the 1-1 component $G_{\,1}^{1}=S_{\,1}^{1}$, that is Eq.~\eqref{subeq: 11 LTB Jordan frame}:
\begin{align}
 & 2\,\ddot{\beta}+3\,\dot{\beta}^{2}+e^{-2\,\beta}-e^{-2\,\alpha}\,\left(\beta^{\prime}\right)^{2}\nonumber \\
 & =\frac{V\left(\phi\right)}{2\,\phi}-\frac{1}{\phi}\,\left[\ddot{\phi}+2\,\dot{\beta}\,\dot{\phi}-2\,e^{-2\,\alpha}\,\beta^{\prime}\,\phi^{\prime}\right]\,.\label{eq: 11 LTB Jordan frame-app}
\end{align}

To verify the isotropy, we start from Eq.~\eqref{eq: 11 LTB Jordan frame-app} and search for a proper relation to obtain  Eq.~\eqref{eq: 22 LTB Jordan frame-app}. Trying to generalize the relation~\eqref{eq:relazioneG11-G22-GR} valid in GR, we can write
\begin{equation}
G_{\,1}^{1}+\frac{1}{2\,\beta^{\prime}}\,\left(G_{\,1}^{1}\right)^{\prime}=S_{\,1}^{1}+\frac{1}{2\,\beta^{\prime}}\,\left(S_{\,1}^{1}\right)^{\prime}\,,\label{eq: prop intermedia}
\end{equation}
which is equivalent to
\begin{align}
 & \ddot{\alpha}+\ddot{\beta}+\dot{\alpha}^{2}+\dot{\beta}\left(\dot{\alpha}+\dot{\beta}\right)-e^{-2\alpha}\left[\beta^{\prime\prime}-\beta^{\prime}\left(\alpha^{\prime}-\beta^{\prime}\right)\right]\nonumber \\
 & \,\,-\frac{\phi^{\prime}}{2\,\phi\,\beta^{\prime}}\,G_{\textrm{extra}}=\frac{V\left(\phi\right)}{2\,\phi}-\frac{\phi^{\prime}}{2\,\phi\,\beta^{\prime}}\,S_{\textrm{extra}}\nonumber \\
 & \,\,-\frac{1}{\phi}\left\{ \ddot{\phi}+\left(\dot{\alpha}+\dot{\beta}\right)\dot{\phi}-e^{-2\alpha}\left[\phi^{\prime\prime}-\phi^{\prime}\left(\alpha^{\prime}-\beta^{\prime}\right)\right]\right\} \,.\label{eq: prop intermedia2}
\end{align}
We have considered geometric $G_{\,1}^{1}$ and source $S_{\,1}^{1}$ contributions in Eq.~\eqref{eq: 11 LTB Jordan frame-app} and their derivatives with respect to $r$, and we have defined
\begin{align}
G_{\text{extra}} &=\frac{\ddot{\phi}^{\prime}}{\phi^{\prime}}-\ddot{\alpha}-\dot{\alpha}^{2}+\left(\frac{\dot{\phi}}{\phi}-2\,\dot{\beta}\right)\,\left(\dot{\alpha}-\frac{\dot{\phi}^{\prime}}{\phi^{\prime}}\right)\label{eq:gextra}\\
S_{\text{extra}} &=\frac{\ddot{\phi}^{\prime}}{\phi^{\prime}}-\frac{\ddot{\phi}}{\phi}+\frac{\dot{\phi}}{\phi}\left(\dot{\alpha}-2\dot{\beta}-\frac{\dot{\phi}^{\prime}}{\phi^{\prime}}\right)+2\dot{\beta}\frac{\dot{\phi}^{\prime}}{\phi^{\prime}}\nonumber \\
&+\frac{V\left(\phi\right)}{2\phi}-\frac{1}{2}\frac{dV}{d\phi}+2e^{-2\alpha}\nonumber\\
& \times\left[\beta^{\prime}\left(\alpha^{\prime}-\beta^{\prime}+\frac{\phi^{\prime}}{\phi}\right)-\beta^{\prime\prime}\right].\label{eq:sextra}
\end{align}

Then, comparing Eqs.~\eqref{eq: 22 LTB Jordan frame-app} and \eqref{eq: prop intermedia2}, it is quite immediate to recognize that the latter equation becomes
\begin{equation}
G_{\,2}^{2}=S_{\,2}^{2}+\frac{\phi^{\prime}}{2\,\phi\,\beta^{\prime}}\,\left(G_{\text{extra}}-S_{\text{extra}}\right)\,.\label{eq: prop intermedia 3}
\end{equation}
Note that, as we previously said, the relation \eqref{eq:relazioneG11-G22-GR} is no longer valid in the Jordan frame since now we have
\begin{equation}
G_{\,2}^{2}=G_{\,1}^{1}+\frac{1}{2\,\beta^{\prime}}\,\left(G_{\,1}^{1}\right)^{\prime}+\frac{\phi^{\prime}}{2\,\phi\,\beta^{\prime}}\,G_{\text{extra}}\,,\label{eq:relazione-G11-G22-JF}
\end{equation}
which is due to an extra term containing the non-minimal coupled scalar field $\phi$.

As a last point, we show that the bracket in Eq.~\eqref{eq: prop intermedia 3} vanishes. Indeed, it is straightforward to identify the difference $\left(G_{\text{extra}}-S_{\text{extra}}\right)=0$, since it exactly coincides with the supplementary Eq.~\eqref{subeq: effect tensor mu=00003D1 LTB} given by the effective stress-energy tensor $T_{\mu\nu}^{[\phi]}$ for $\mu=1$, which is just originated from the gravitational field equations. 

Hence, we have proved that the 2-2 component~\eqref{eq: 22 LTB Jordan frame-app} of the field equations in the Jordan frame depends on the 1-1 component~\eqref{eq: 11 LTB Jordan frame-app} and the additional Eq.~\eqref{subeq: effect tensor mu=00003D1 LTB}. As a consequence, despite the presence of extra coupling terms between the scalar field and the metric, the field equations in the Jordan frame of $f\left(R\right)$ gravity implemented at the LTB metric preserve the spatial isotropy, as it must be for a spherically symmetric solution.

\section{}\label{app:technical}

In this appendix, we want to rewrite the equation system given by the linearly perturbed field equations~\eqref{eq-linearized-JF-LTB}, \eqref{eq:linearized-scalar-field-LTB-JF}, and \eqref{eq:linearized-contin-LTB-JF} to study separately time and space evolution of perturbations through the separation of variables method. The final aim is to show explicit calculations to obtain the two sets of equations \eqref{eq-linearized-JF-LTB-separated-only-r} and \eqref{eq-linearized-JF-LTB-separated-only-t}, which have been reported in \prettyref{subsec:Linear-solution-JF}.

We adopt the factorization~\eqref{eq:separation-variables-LTB-JF} for all linear perturbations, which we rewrite here for convenience:
\begin{align}
& \delta\alpha\left(t,r\right) \equiv A_{p}\left(t\right)\,\mathcal{A}_{p}\left(r\right)\nonumber\\
& \delta\beta\left(t,r\right)\equiv B_{p}\left(t\right)\,\mathcal{B}_{p}\left(r\right)\nonumber \\
& \delta\rho\left(t,r\right) \equiv P_{p}\left(t\right)\,\varrho_{p}\left(r\right)\nonumber\\
& \delta\phi\left(t,r\right)\equiv\Phi_{p}\left(t\right)\,\varphi_{p}\left(r\right)\,.\label{eq:separation-variables-LTB-JF-app}
\end{align}
We have split time and radial dependences for each scalar function.

Then, considering the factorization~\eqref{eq:separation-variables-LTB-JF-app}, the equation system~\eqref{eq-linearized-JF-LTB} becomes:
\begin{subequations}
\label{eq-linearized-JF-LTB-split-t-r}
\begin{align}
 & \frac{\mathcal{B}_{p}^{\prime}}{\varphi_{p}^{\prime}}+\frac{1}{r}\,\frac{\mathcal{B}_{p}}{\varphi_{p}^{\prime}}=\frac{1}{r}\,\frac{\mathcal{A}_{p}}{\varphi_{p}^{\prime}}\,\frac{\dot{A}_{p}}{\dot{B}_{p}}-\frac{1}{2\,\bar{\phi}\,\dot{B}_{p}}\,\left(\dot{\Phi}_{p}-\frac{\dot{\bar{a}}}{\bar{a}}\,\Phi_{p}\right)\,,\label{subeq:linearized-01-LTB-JF-split-t-r}\\
 & \frac{2}{\bar{a}^{2}}\left[\frac{A_{p}}{r}\left(\frac{\mathcal{A}_{p}}{r}+\mathcal{A}_{p}^{\prime}\right)-B_{p}\left(\mathcal{B}_{p}^{\prime\prime}+\frac{3}{r}\mathcal{B}_{p}^{\prime}+\frac{\mathcal{B}_{p}}{r^{2}}\right)\right]\nonumber \\
 & \,\,+\left(2\,\frac{\dot{\bar{a}}}{\bar{a}}+\frac{\dot{\bar{\phi}}}{\bar{\phi}}\right)\,\left(\dot{A}_{p}\,\mathcal{A}_{p}+2\,\dot{B}_{p}\,\mathcal{B}_{p}\right)=\frac{\chi}{\bar{\phi}}\,P_{p}\,\varrho_{p}\nonumber \\
 & \,\,+\left(\frac{1}{2}\left.\frac{dV}{d\phi}\right|_{\phi=\bar{\phi}}-\frac{1}{2}\frac{V\left(\bar{\phi}\right)}{\bar{\phi}}-\frac{\chi\bar{\rho}}{\bar{\phi}}+3\frac{\dot{\bar{a}}}{\bar{a}}\frac{\dot{\bar{\phi}}}{\bar{\phi}}\right)\frac{\Phi_{p}\varphi_{p}}{\bar{\phi}}\nonumber \\
 & \,\,+\frac{1}{\bar{a}^{2}\,\bar{\phi}}\,\left(\Phi_{p}\,\varphi_{p}^{\prime\prime}+\frac{2}{r}\,\Phi_{p}\,\varphi_{p}^{\prime}\right)-3\,\frac{\dot{\bar{a}}}{\bar{a}}\,\frac{\dot{\Phi}_{p}\,\varphi_{p}}{\bar{\phi}}\,,\label{subeq:linearized-00-LTB-JF-split-t-r}\\
 & \left(\ddot{B}_{p}+3\frac{\dot{\bar{a}}}{\bar{a}}\dot{B}_{p}\right)\mathcal{B}_{p}+\frac{1}{\bar{a}^{2}r^{2}}\left[A_{p}\mathcal{A}_{p}-B_{p}\left(\mathcal{B}_{p}+r\mathcal{B}_{p}^{\prime}\right)\right]\nonumber \\
 & =\left(\frac{1}{2}\,\left.\frac{dV}{d\phi}\right|_{\phi=\bar{\phi}}-\frac{V\left(\bar{\phi}\right)}{2\,\bar{\phi}}+\frac{\ddot{\bar{\phi}}}{\bar{\phi}}+2\,\frac{\dot{\bar{a}}}{\bar{a}}\,\frac{\dot{\bar{\phi}}}{\bar{\phi}}\right)\,\frac{\Phi_{p}\,\varphi_{p}}{2\,\bar{\phi}}\nonumber \\
 & \,-\frac{1}{\bar{\phi}}\,\left(\frac{1}{2}\,\ddot{\Phi}_{p}\,\varphi_{p}+\frac{\dot{\bar{a}}}{\bar{a}}\,\dot{\Phi}_{p}\,\varphi_{p}+\dot{\bar{\phi}}\,\dot{B}_{p}\,\mathcal{B}_{p}-\frac{\Phi_{p}\,\varphi_{p}^{\prime}}{\bar{a}^{2}\,r}\right)\,.\label{subeq:linearized-11-LTB-JF-split-t-r}
\end{align}
\end{subequations}

We have tried to separate time-dependent and radial terms. However, this equations system exhibits a complicated structure, since it should be noted the occurrence of several mixed terms depending both on $t$ and $r$. For instance, focusing on Eq.~\eqref{subeq:linearized-01-LTB-JF-split-t-r}, terms on the left-hand side depend only on $r$, a mixed term is the first contribution on the right side, while the second contribution is only time-dependent. 

Using again Eq.~\eqref{eq:separation-variables-LTB-JF-app}, the linearized scalar field equation~\eqref{eq:linearized-scalar-field-LTB-JF} rewrites as
\begin{align}
 & \bar{a}^{2}\,\left[\frac{\ddot{\Phi}_{p}}{\Phi_{p}}+3\,\frac{\dot{\bar{a}}}{\bar{a}}\,\frac{\dot{\Phi}_{p}}{\Phi_{p}}+\left(\frac{\dot{A}_{p}}{\Phi_{p}}\,\frac{\mathcal{A}_{p}}{\varphi_{p}}+2\,\frac{\dot{B}_{p}}{\Phi_{p}}\,\frac{\mathcal{B}_{p}}{\varphi_{p}}\right)\,\dot{\bar{\phi}}\right.\nonumber \\
 & \quad\left.-\frac{1}{3}\,\left.\frac{dV}{d\phi}\right|_{\phi=\bar{\phi}}+\frac{1}{3}\,\bar{\phi}\,\left.\frac{d^{2}V}{d\phi^{2}}\right|_{\phi=\bar{\phi}}-\frac{\chi}{3}\,\frac{P_{p}}{\Phi_{p}}\,\frac{\varrho_{p}}{\varphi_{p}}\right]=\nonumber \\
 & \quad\frac{\varphi_{p}^{\prime\prime}}{\varphi_{p}}\,+\frac{2}{r}\,\frac{\varphi_{p}^{\prime}}{\varphi_{p}}\,,\label{eq:linearized-scalar-field-LTB-JF-split-t-r}
\end{align}
while the linearized continuity equation~\eqref{eq:linearized-contin-LTB-JF} becomes
\begin{equation}
\frac{1}{\bar{\rho}\,\dot{B}_{p}}\,\left(\dot{P}_{p}+3\,\frac{\dot{\bar{a}}}{\bar{a}}\,P_{p}\right)+\frac{\dot{A}_{p}}{\dot{B}_{p}}\,\frac{\mathcal{A}_{p}}{\varrho_{p}}+2\,\frac{\mathcal{B}_{p}}{\varrho_{p}}=0\,.\label{eq:linearized-contin-LTB-JF-split-t-r}
\end{equation}

We noticed again the presence of mixed terms depending on $t$ and $r$, which do not allow us to solve the equations using the separation of variables unless we rely on some simplifying assumptions. For instance, focusing on Eqs.~\eqref{subeq:linearized-01-LTB-JF-split-t-r} and \eqref{eq:linearized-contin-LTB-JF-split-t-r}, if we require that the perturbations $A_{p}$ and $B_{p}$ follow a similar time evolution, i.e.
\begin{equation}
\dot{A}_{p}=\lambda_{1}\,\dot{B}_{p}\,,
\end{equation}
where $\lambda_{1}$ is a constant, then we are able to solve these two equations through the separation of variables method. Actually, we also write equivalently
\begin{equation}
A_{p}=\lambda_{1}\,B_{p}\,,\label{eq:ipotesi-semplificativa1-app}
\end{equation}
which is just Eq.~\eqref{subeq:ipotesi-semplificativa1}, since we can adjust constant term in the first-order perturbation theory. 

Then, if we impose the assumption~\eqref{eq:ipotesi-semplificativa1-app} in Eq.~\eqref{subeq:linearized-01-LTB-JF-split-t-r}, we obtain two equations, one in the variable $t$ and the other one in $r$:
\begin{subequations}
\label{eq-linearized-01-JF-LTB-separated-t-r}
\begin{align}
\dot{B}_{p} & =\frac{1}{2\,\bar{\phi}\,\mu_{1}}\,\left(\dot{\Phi}_{p}-\frac{\dot{\bar{a}}}{\bar{a}}\,\Phi_{p}\right)\,,\label{subeq:linearized-01-LTB-JF-split-t-app}\\
\mathcal{A}_{p} & =\frac{1}{\lambda_{1}}\,\left[\mathcal{B}_{p}+r\,\left(\mathcal{B}_{p}^{\prime}+\mu_{1}\,\varphi_{p}^{\prime}\right)\right]\,,\label{subeq:linearized-01-LTB-JF-split-r-app}
\end{align}
\end{subequations}
which are just Eqs.~\eqref{subeq:linearized-01-LTB-JF-split-t} and \eqref{subeq:linearized-01-LTB-JF-split-r}, respectively.

In the same way, we can split Eq.~\eqref{eq:linearized-contin-LTB-JF-split-t-r} into Eqs.~\eqref{subeq:linearized-contin-LTB-JF-split-t} and \eqref{subeq:linearized-contin-LTB-JF-split-r}
\begin{subequations}
\label{eq-linearized-contin-JF-LTB-separated-t-r}
\begin{align}
 & \dot{P}_{p}+3\,\frac{\dot{\bar{a}}}{\bar{a}}\,P_{p}+\mu_{2}\,\bar{\rho}\,\dot{B}_{p}=0\,,\label{subeq:linearized-contin-LTB-JF-split-t-app}\\
 & \varrho_{p}=\frac{1}{\mu_{2}}\,\left(\lambda_{1}\,\mathcal{A}_{p}+2\,\mathcal{B}_{p}\right)\,,\label{subeq:linearized-contin-LTB-JF-split-r-app}
\end{align}
\end{subequations}
respectively, where $\mu_{1}$ and $\mu_{2}$ are constants originating from the separation of variables. 

Concerning the linearized scalar field equation~\eqref{eq:linearized-scalar-field-LTB-JF-split-t-r}, using Eqs.~\eqref{eq:ipotesi-semplificativa1-app} and \eqref{subeq:linearized-contin-LTB-JF-split-r-app}, we end up in
\begin{align}
 & \bar{a}^{2}\,\left[\frac{\ddot{\Phi}_{p}}{\Phi_{p}}+3\,\frac{\dot{\bar{a}}}{\bar{a}}\,\frac{\dot{\Phi}_{p}}{\Phi_{p}}-\frac{1}{3}\,\left.\frac{dV}{d\phi}\right|_{\phi=\bar{\phi}}+\frac{1}{3}\,\bar{\phi}\,\left.\frac{d^{2}V}{d\phi^{2}}\right|_{\phi=\bar{\phi}}\right.\nonumber \\
 & \quad\left.+\left(\mu_{2}\,\dot{\bar{\phi}}\,\frac{\dot{B}_{p}}{\Phi_{p}}-\frac{\chi}{3}\,\frac{P_{p}}{\Phi_{p}}\right)\,\frac{\varrho_{p}}{\varphi_{p}}\right]=\frac{\varphi_{p}^{\prime\prime}}{\varphi_{p}}\,+\frac{2}{r}\,\frac{\varphi_{p}^{\prime}}{\varphi_{p}}\,.\label{eq:eq:linearized-scalar-field-intermediate-LTB-JF-split-t-r}
\end{align}

At this point, noting a mixed term in the last contribution of the left-hand side, to proceed analytically with the separation of variables, we require an additional simplifying assumption, that is the proportionality between the radial evolution of the matter and scalar field perturbations:
\begin{equation}
\varrho_{p}=\lambda_{2}\,\varphi_{p}\,,\label{eq:ipotesi-semplificativa2-app}
\end{equation}
i.e., Eq.~\eqref{subeq:ipotesi-semplificativa2}, where $\lambda_{2}$ is the proportionality constant. 

Hence, we can easily separate time and radial evolutions in Eq.~\eqref{eq:eq:linearized-scalar-field-intermediate-LTB-JF-split-t-r} to write the two differential equations \eqref{subeq:linearized-scalar-field-LTB-JF-split-t} and \eqref{subeq:linearized-scalar-field-LTB-JF-split-r}, respectively:
\begin{subequations}
\label{eq-linearized-scalar-field-JF-LTB-separated-t-r}
\begin{align}
 & \,\frac{\ddot{\Phi}_{p}}{\Phi_{p}}+3\,\frac{\dot{\bar{a}}}{\bar{a}}\,\frac{\dot{\Phi}_{p}}{\Phi_{p}}-\frac{1}{3}\,\left.\frac{dV}{d\phi}\right|_{\phi=\bar{\phi}}+\frac{1}{3}\,\bar{\phi}\,\left.\frac{d^{2}V}{d\phi^{2}}\right|_{\phi=\bar{\phi}}\nonumber \\
 & \quad+\mu_{2}\,\lambda_{2}\,\dot{\bar{\phi}}\,\frac{\dot{B}_{p}}{\Phi_{p}}-\frac{\chi\,\lambda_{2}}{3}\,\frac{P_{p}}{\Phi_{p}}=\frac{\mu_{3}^{2}}{\bar{a}^{2}}\,,\label{subeq:linearized-scalar-field-LTB-JF-split-t-app}\\
 & \varphi_{p}^{\prime\prime}+\frac{2}{r}\,\varphi_{p}^{\prime}-\mu_{3}^{2}\,\varphi_{p}=0\,,\label{subeq:linearized-scalar-field-LTB-JF-split-r-app}
\end{align}
\end{subequations}
where $\mu_{3}$ is a constant. 

If we focus on Eq.~\eqref{subeq:linearized-00-LTB-JF-split-t-r}, by simplifying mixed terms with Eqs.~\eqref{eq:ipotesi-semplificativa1-app}, \eqref{subeq:linearized-01-LTB-JF-split-t-app}, \eqref{subeq:linearized-01-LTB-JF-split-r-app}, \eqref{eq:ipotesi-semplificativa2-app}, and \eqref{subeq:linearized-contin-LTB-JF-split-r-app}, it is straightforward to show that, after long calculations, we encompass the radial part through Eq.~\eqref{subeq:linearized-scalar-field-LTB-JF-split-r-app}, and we obtain a single equation in $t$:
\begin{align}
 & \frac{\mu_{2}\lambda_{2}}{2\mu_{1}}\left(2\frac{\dot{\bar{a}}}{\bar{a}}+\frac{\dot{\bar{\phi}}}{\bar{\phi}}\right)\left(\frac{\dot{\Phi}_{p}}{\Phi_{p}}-\frac{\dot{\bar{a}}}{\bar{a}}\right)+\frac{\chi\bar{\rho}}{\bar{\phi}}-\frac{1}{2}\left.\frac{dV}{d\phi}\right|_{\phi=\bar{\phi}}\nonumber \\
 & \,\,+\frac{V\left(\bar{\phi}\right)}{2\bar{\phi}}+3\frac{\dot{\bar{a}}}{\bar{a}}\left(\frac{\dot{\Phi}_{p}}{\Phi_{p}}-\frac{\dot{\bar{\phi}}}{\bar{\phi}}\right)-\chi\lambda_{2}\frac{P_{p}}{\Phi_{p}}\nonumber\\
 & \,\,+2\mu_{1}\mu_{3}^{2}\frac{\bar{\phi}}{\bar{a}^{2}}\frac{B_{p}}{\Phi_{p}}=\frac{\mu_{3}^{2}}{\bar{a}^{2}}\,,\label{eq:00-LTB-JF-split-t-app}
\end{align}
which is just Eq.~\eqref{subeq:linearized-00-LTB-JF-split-t}.

Finally, regarding the last equation of the system \eqref{eq-linearized-JF-LTB-split-t-r} to be rewritten with the separation of variables, that is Eq.~\eqref{subeq:linearized-11-LTB-JF-split-t-r}, if we combine it with Eqs.~\eqref{eq:ipotesi-semplificativa1-app}, \eqref{subeq:linearized-01-LTB-JF-split-t-app}, \eqref{subeq:linearized-01-LTB-JF-split-r-app}, we obtain
\begin{subequations}
\label{eq-linearized-11-JF-LTB-separated-t-r}
\begin{align}
 & \ddot{B}_{p}+\dot{B}_{p}\,\left(3\,\frac{\dot{\bar{a}}}{\bar{a}}+\frac{\dot{\bar{\phi}}}{\bar{\phi}}\right)=\frac{\mu_{4}}{2\,\bar{a}^{2}}\,\left(\frac{\Phi_{p}}{\bar{\phi}}-\mu_{1}\,B_{p}\right)\,,\label{subeq:linearized-11-LTB-JF-split-t-app}\\
 & \mathcal{B}_{p}=\frac{2}{\mu_{4}\,r}\,\varphi_{p}^{\prime}-\mu_{1}\,\varphi_{p}\,,\label{subeq:linearized-11-LTB-JF-split-r-app}
\end{align}
\end{subequations}
which are exactly Eqs.~\eqref{subeq:linearized-11-LTB-JF-split-t} and \eqref{subeq:linearized-11-LTB-JF-split-r}, respectively, where $\mu_{4}$ is a constant.

In conclusion, in this appendix, we have shown how the two simplifying assumptions~\eqref{eq:ipotesi-semplificativa1-app} and \eqref{eq:ipotesi-semplificativa2-app} are suggested from the analysis of the equation system given by Eqs.~\eqref{eq-linearized-JF-LTB}, \eqref{eq:linearized-scalar-field-LTB-JF}, and \eqref{eq:linearized-contin-LTB-JF} to use the separation of variables. Finally, we have obtained separately two sets of equations~\eqref{eq-linearized-JF-LTB-separated-only-r} and \eqref{eq-linearized-JF-LTB-separated-only-t} for the radial and time evolution of linear perturbations, which have been reported in \prettyref{subsec:Linear-solution-JF}. Once we have split all field equations into time and space components, we can solve them to obtain linear order perturbations separately in \prettyref{subsec:Radial-solutions-JF} and \prettyref{subsec:Time-solutions-JF}.

\end{appendices}

\bibliography{articleLTB}

\end{document}